\documentclass[%
superscriptaddress,
amsmath,amssymb,
aps,color,
twocolumn,
prl,
]{revtex4-2}

\usepackage{graphicx}
\usepackage{dcolumn}
\usepackage{bm}
\usepackage{hyperref}
\usepackage{siunitx}
\usepackage{physics}   
\usepackage{ulem}
\usepackage{booktabs}
\usepackage{orcidlink}
\usepackage{enumitem}

\usepackage{booktabs}
\setenumerate{label={\bf{\arabic{enumi}}.}, itemsep=3pt}
\setitemize{label={\raisebox{1pt}{\tiny\textbullet}}, itemsep= 1pt}

\usepackage{breqn}
\usepackage{etoolbox}
\makeatletter
\preto\maketitle{%
	\begingroup\lccode`~=`,
	\lowercase{\endgroup
		\let\saved@breqn@active@comma~
		\let~}\active@comma 
}
\appto\maketitle{%
	\begingroup\lccode`~=`,
	\lowercase{\endgroup
		\let~}\saved@breqn@active@comma 
}
\makeatother

\begin{document}
	\newcommand{\rr}[1]{\textcolor{red}{[#1]}}

	\preprint{APS/123-QED}
	
	\title{Controlling confined collective organisation with taxis}
	
	\author{Albane Th\'ery    \orcidlink{0000-0001-7465-4849}}
	\email{athery@sas.upenn.edu}
	\affiliation{Department of Applied Mathematics and Theoretical Physics, University of Cambridge,	
		Cambridge CB3 0WA, UK}
  \affiliation{Department of Mathematics, University of Pennsylvania,	
		19104 Philadelphia PA, US}
	\author{Alexander Chamolly \orcidlink{0000-0002-2383-9314}}
	\email{alexander.chamolly@pasteur.fr}
	\affiliation{Institut Pasteur, Universit\'e de Paris, CNRS UMR3738, Developmental and Stem Cell Biology Department, F-75015 Paris, France}
	\affiliation{
		Laboratoire de Physique de l’Ecole normale sup\'erieure, ENS, Universit\'e PSL, CNRS, Sorbonne
		Universit\'e, Universit\'e de Paris, F-75005 Paris, France 
	}
	\author{Eric Lauga \orcidlink{0000-0002-8916-2545}}
	\email{e.lauga@damtp.cam.ac.uk}
	\affiliation{Department of Applied Mathematics and Theoretical Physics, University of Cambridge,	
		Cambridge CB3 0WA, UK}

	\date{\today}
	
	\begin{abstract}
Biased locomotion is a common feature of microorganisms, but little is known about its impact on self-organisation. Inspired by recent experiments showing a transition to large-scale flows, we study theoretically the dynamics of magnetotactic bacteria confined to a drop. We reveal two symmetry-breaking mechanisms (one local chiral and one global achiral) leading to self-organisation into global vortices and a net torque exerted on the drop. The collective behaviour is ultimately controlled by the swimmers' microscopic chirality and, strikingly, the system can exhibit oscillations and memory-like features.

 \end{abstract}

\maketitle



\begin{figure}[b!]
\centering
	\includegraphics[width = \columnwidth]{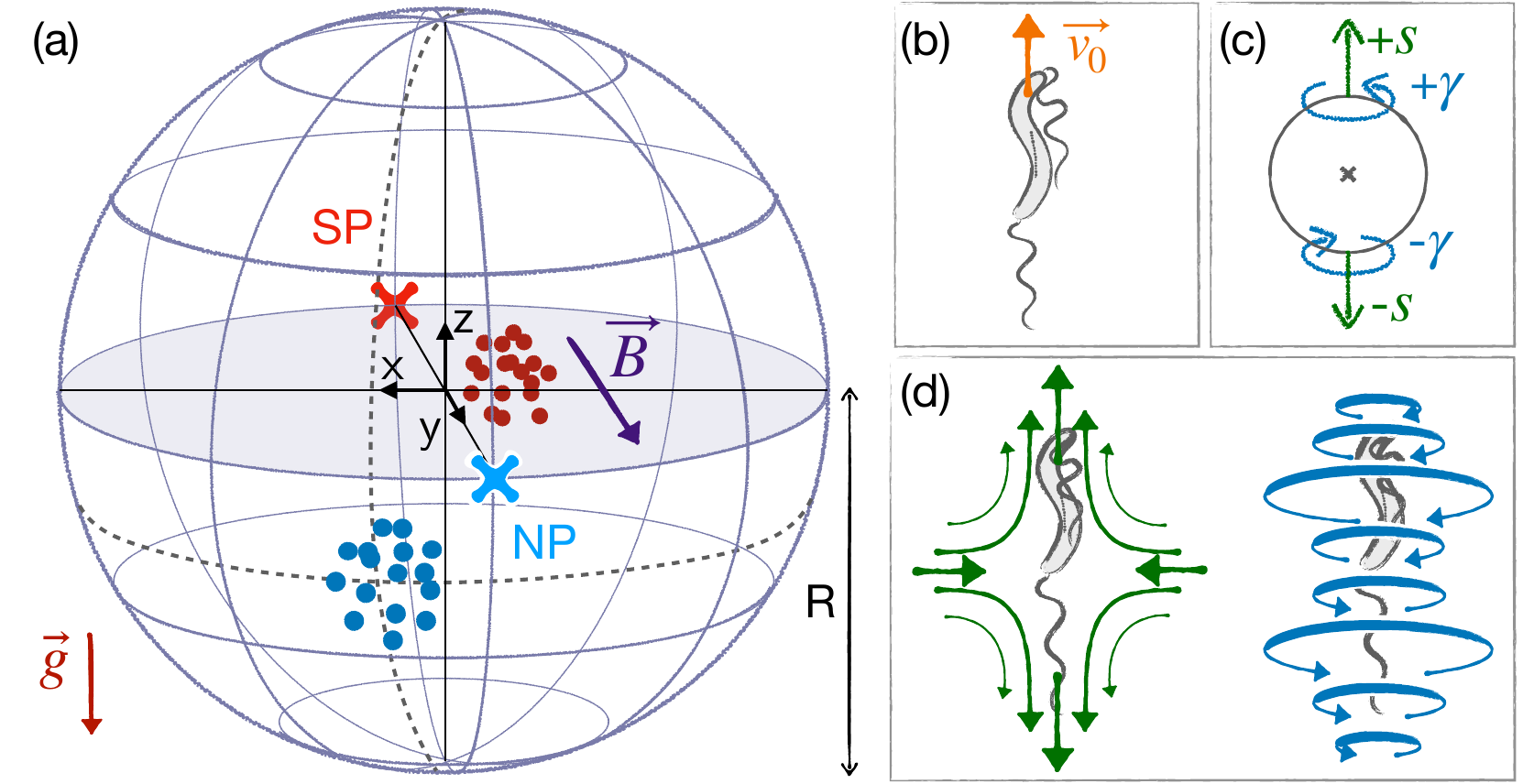}
	\caption[] {\label{fig1} (a) Magnetotactic bacteria accumulate at the North (NP) and South Pole (SP) of the spherical drop. (b) Sketches of an individual MTB, (c) its hydrodynamic singularity model as dipoles of forces ($s$) and torques ($\gamma$), and (d) corresponding flow fields.
}
\end{figure}

As it is for their macroscopic counterparts, taxis -- the ability to move in response to environmental cues -- is crucial to the life of motile microorganisms~\cite{braybook}; they can respond to gradients in chemicals~\cite{soto2014self}, light~\cite{vincent1996bioconvection}, viscosity~\cite{Liebchen2018viscotaxis} or gravity~\cite{pedley_2015}. In turn, the directed motion of individuals drives new collective dynamics, such as the well-known bioconvection of gyrotactic or phototactic algae \cite{vincent1996bioconvection, Hill_2005} or clustering instabilities~\cite{Lauga_2016}. A bias in the swimmer dynamics can take the form of passive control of cell orientation. For example, magnetotactic bacteria (MTB) exhibit magnetic moments and align to external fields~\cite{Blakemore_1975}. In suspensions, this leads to pearling instabilities~\cite{waisbord2016destabilization,Meng2018}, boundary-mediated clustering~\cite{Petroff_2015,pierce2017tuning,Pierce_2020} and plume formation~\cite{Thery_2020}. 
A consistent feature of biased collective dynamics is the prominence of confinement-mediated interactions; oriented swimmers tend to accumulate at boundaries and create dense regions where self-organisation occurs~\cite{pedley1992hydrodynamic, hill2005bioconvection, Thery_2020}. Given its significant potential for biological and biomedical control, a predictive framework for such self-organisation is needed.
Striking recent experiments showed that 
MTB can set in motion droplets 40 times larger than individual cells~\cite{vincenti2019magnetotactic}. Specifically, suspensions of MTB inside water-in-oil droplets under a horizontal external magnetic field $\mathbf{B}=B \mathbf{\hat{y}}$  (Fig.~\ref{fig1}a) spontaneously self-organise in a single large-scale vortex, thereby rotating the whole drop around the vertical $z$-axis. 
While global vortical flows are a staple of collective motion in circular or spherical confinement without director fields~\cite{Opathalage_2019,Chen_2021,Huang_2021}, in particular for bacteria~\cite{lushi2014fluid,wioland2013confinement,Vladescu2014,Hamby_2018,Creppy_2016}, the vortex rotation direction is usually random~\cite{Tsang_2015}. 
Not so for MTB, where the vortex is consistently oriented in the positive $z$-direction~\cite{vincenti2019magnetotactic}, termed clockwise (CW), irrespective of the sign of $B$, hinting at a different mechanism for the onset of large-scale motion. 
Surprisingly, the vortex reverses to a negative $z$-rotation when the field is reversed to $-B \hat{\mathbf{y}}$ \cite{vincenti2019magnetotactic}: a seemingly identical system now exhibits a different, counter-clockwise (CCW) bias. The collective behaviour therefore depends on the history of actuation, despite an underlying linear and inertialess flow.

Here, we uncover the path to symmetry breaking and vortex formation in suspensions of biased swimmers under spherical confinement. Using simulations featuring long-range hydrodynamic interactions between cells with a chiral propulsive component, steric repulsion, and sedimentation, we identify two competing mechanisms for the onset of global rotation in the suspension. 
One, denoted (G), is global and stems from the interaction between two achiral populations at opposite poles of the drop,  while the second is local, (L), and displayed by a single population of chiral swimmers gathering at one pole. 
Both mechanisms are explained through a minimal mathematical model involving hydrodynamic singularities. Notably, the local mechanism has a preferred rotation direction stemming from the swimmers' chirality and leads to systematic CW rotation, as in Ref.~\cite{vincenti2019magnetotactic}. This symmetry-breaking stems from the interaction of biased swimmers with an inclined (droplet) surface and is expected to be observed in other, non-spherical geometries. 
We also predict the onset of oscillatory flows for stronger, but physically relevant, chiralities. Moreover, the subtle competition between the two mechanisms controls the flow direction upon field reversal. Because of its sensitivity to chirality, this setup can be used to estimate and compare the chirality of biased microswimmers, a characteristic otherwise difficult to evaluate experimentally.


\paragraph{Minimal model.} We show that symmetry breaking is a physical rather than biological phenomenon with a minimal model for a suspension of $N_s$ biased swimmers (Fig.~\ref{fig1}), set up as follows. 
Each swimmer is described by its position and orientation. Initially, they are homogeneously distributed with random orientations within a sphere of radius $R$ filled with a fluid of viscosity $\mu$. 
Their dynamics are then set by (i) their alignment to the external magnetic field, (ii) self-propulsion, (iii) sedimentation, (iv) noise, and (v) steric and (vi) hydrodynamic interactions with other swimmers and the droplet boundary. 

In detail, (i) a globally preferred orientation is set by $B\hat{\mathbf{y}}$, 
 taking the role of an external director mechanism. The suspension is evenly split into north (NS) and south-seeking (SS) populations, which experience alignment torques towards the North (NP) and South pole (SP) respectively (Fig.~\ref{fig1}a). The alignment strength is set by the field $B$ and the magnetic moment normalised by the rotational drag coefficient $\tilde m$. 
The bacteria also (ii) swim at a constant speed $v_0$ and (iii) sediment due to a slight density difference with the medium ($-v_g\hat{\mathbf{z}}$). 
(iv) Thermal and active noise are included through Brownian noise in translation (diffusivity $D_t$) and rotation ($D_r$). 
(v) Hard-sphere steric interactions enforce a repulsion between nearby swimmers and keep them inside the sphere. 
(vi) Hydrodynamic interactions occur with the droplet surface, as well as with the flow from surrounding swimmers, themselves modified by the presence of a boundary.

In the dilute limit, hydrodynamic interactions are governed by the leading-order flow signature of the magnetotactic bacteria MSR-1 (Fig.~\ref{fig1}d), which is an extensile force dipole (strength $s$) since cells are pushed by their aft helical flagella~\cite{lauga2009hydrodynamics,Reufer_2014}. 
Additionally, the flagellar rotation and associated body counter-rotation induce localised hydrodynamic torques, modelled as a rotlet dipole (strength $\gamma$),  which decays faster than the stresslet but is the leading-order chiral signature. To explain the symmetry breaking,   both flow singularities turn out to be necessary. 
To enforce the no-slip boundary condition on the droplet surface exactly, we use the method of images. We incorporate a perturbation to the bulk flow resulting from additional hydrodynamic singularities outside the sphere to match the flow due to the force~\cite{maul1994image,shail1987note} and torque~\cite{hackborn1986structure,shail1987note} dipoles on the boundary~\cite{chamolly2020stokes}. 
Each individual swimmer is then advected and reoriented by the flow from its own image and the sum of bulk and image flows from other cells.

We scale lengths by the cell size $d$, time by $d/v_0$, forces by $\mu v_0 d$ and magnetic field strengths by $v_0/(\tilde{m} d)$, and use dimensionless quantities; 
setup parameters are taken from Ref~\cite{vincenti2019magnetotactic}. Values specific to the bacteria are drawn from experiments with MSR-1 when available~\cite{Reufer_2014, Zahn_2017, Waisbord_2021} and we substitute unknown values of $s$ and $\gamma$ by those for \textit{E.~coli}~\cite{Ford_2007,Hu_2015}. The supplementary material contains details of the numerical integration and the influence of all parameters.



\begin{figure}[t!]
    \includegraphics[width =\columnwidth]{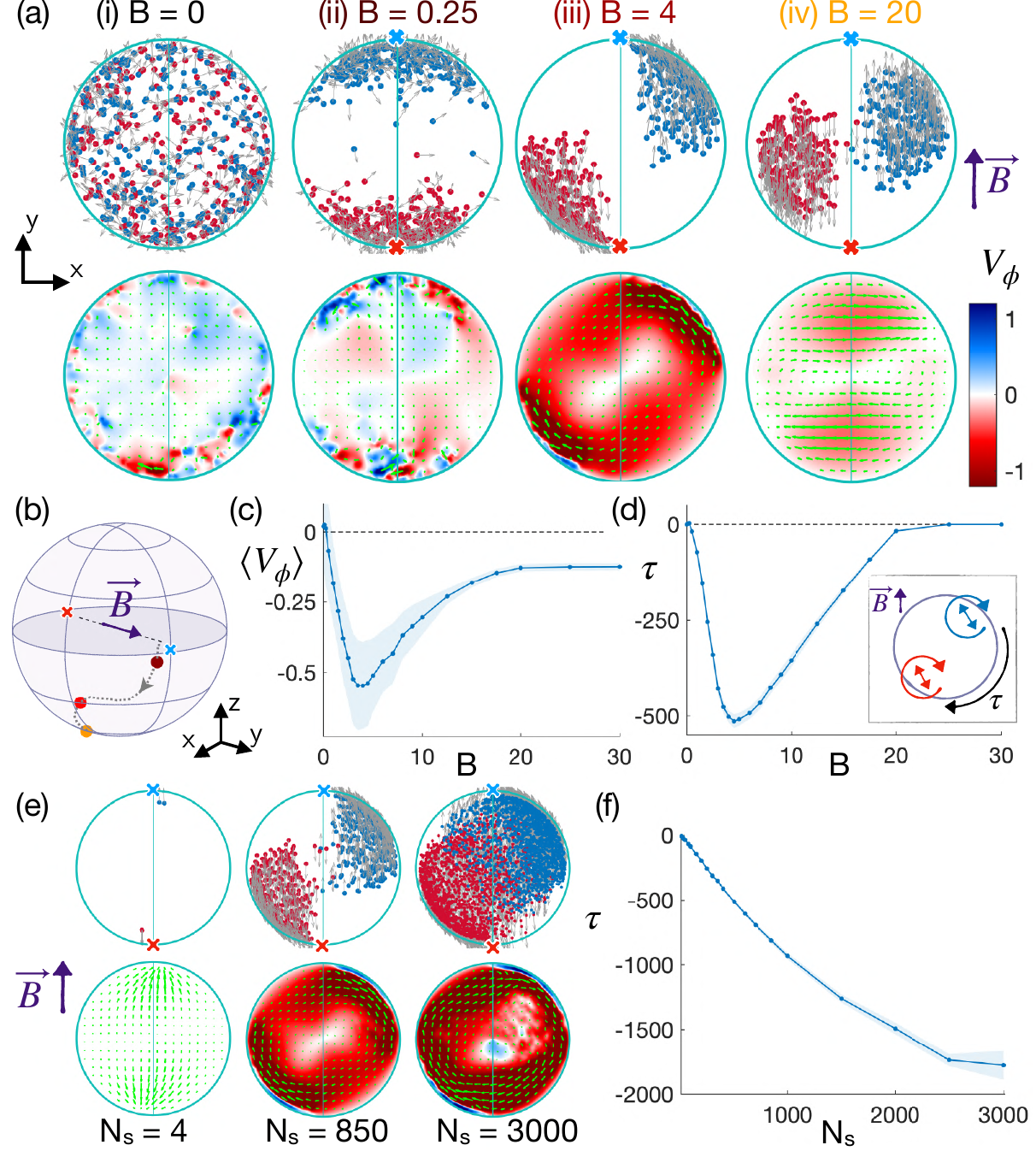}
    \caption{\label{figpar} symmetry breaking and flow transition under an external field. 
    (a) Snapshots for different magnetic field strengths. 
    Top: position of the NS (blue) and SS (red) swimmers. Bottom: flow in the equatorial plane, in-plane velocity vectors (green), superposed with a density plot of the azimuthal component, $V_\phi$.
    (i) Disordered system at $B=0$. 
    (ii) Clusters form at the poles when the field is turned on. (iii, iv) Clusters slide CW at higher fields, creating a global CW vortex in the $x$-$y$ plane. 
    (b) Average position of the above NS clusters. 
  (c) Mean value of spatially averaged vortex strength, $\langle V_\phi \rangle$, with shaded steady-state standard deviation. The CW vortex is strongest at $B=B^* \sim 4$. 
    (d) Total magnetic torque, $\tau$, as a function of $B$. Inset: sketch of the torque generation mechanism, which vanishes when $B=0$ or $B > 20$. 
    (e) Phenomenology for different concentrations: ($N_s=4$) Broken symmetry for just two swimmers from each population, ($N_s=850$) emergence of a vortex, and ($N_s=3000$) partial mixing at high concentrations 
    (f) Torque on the drop as a function of $N_s$.}
\end{figure}

\begin{figure*}[t!]
\centering
		\includegraphics[ width = 175 mm]{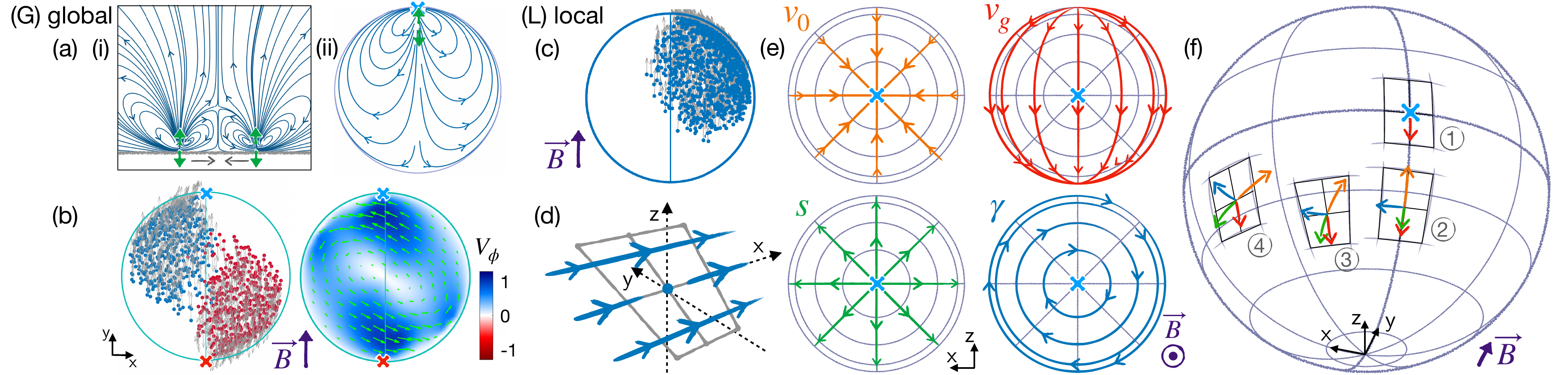}
		\caption{\label{figmec} Global (G) and Local (L) mechanisms for symmetry breaking. 
		(a) (i) Streamlines from two force dipoles perpendicular to a wall. The inward lateral flow results in an effective attraction (grey arrows). (ii) Streamlines from a force dipole at the NP. For a cluster at the SP, this flow is destabilising. 
    (b) (G)-symmetry breaking for two achiral ($\gamma = 0$ populations, with no preferred CW/CCW direction. 		
    (c) Directed (L) CW symmetry breaking for a single chiral population. 
    (d) Net flow from a dipole of torques $\gamma$ along $\mathbf{\hat{y}}$ 
    near a no-slip inclined boundary.
    (e) Contributions to the NS cluster displacement: self-propulsion $v_0$, gravity $v_g$, force dipoles $s$ (attraction to parallel walls), torque dipoles $\gamma$ (CW precession around the NP). 
    (f) Dynamic sequence of cluster positions during a (L) symmetry breaking, from the NP (1) to the steady-state (4).  }
\end{figure*}

\paragraph{Results.} Our simulations reveal that the magnetic field strength, $B$, acts as the control parameter for a transition from disorder without director field 
to a self-organised state where the $\pm x$-symmetry is broken (Fig.~\ref{figpar}a). 
For weak fields, the NS and SS cell populations separate and accumulate at their preferred pole, but the system remains symmetric in $\pm x$ (Fig.~\ref{figpar}a(ii)). 
As $|B|$ increases, the clusters move azimuthally to an asymmetric steady-state configuration, displaced clockwise from the magnetic poles (Fig.~\ref{figpar}a(iii)-(iv) and b). As a result of this shift in position, a global CW vortex emerges in the $x$-$y$ plane, 
accompanied by two narrow recirculation regions between each cluster and the boundary, in agreement with experimental observations \cite{vincenti2019magnetotactic}.

The average azimuthal component of the fluid velocity in the equatorial plane, $\langle V_\phi \rangle$ (Fig.~\ref{figpar}c), quantifies the transition to a vortical flow. It exhibits a maximum at a strong but intermediate value of the dimensionless field $B^* \sim 4$ (above the range of Ref.~\cite{vincenti2019magnetotactic}). 
This optimum stems from a competition between alignment with the boundary and with the field. 
Partial alignment with the boundary at weak fields (Fig.~\ref{figpar}a(iii)) promotes upward swimming against gravity and a greater separation between clusters, leading to stronger circulation. At higher field strengths, symmetry is still broken but the vortex generated by the sedimented swimmers is weaker (Fig.~\ref{figpar}a(iv)). 
The misalignment of swimmers with the field at the boundary also results in a net magnetic torque which is predominantly CW in both NS and SS cell clusters (Fig.~\ref{figpar}d, inset). This net torque ($\tau$) is transmitted to the droplet, which rotates if freely suspended, as seen experimentally~\cite{vincenti2019magnetotactic}, with an optimum close to $B^*$ (Fig.~\ref{figpar}d).

Notably, symmetry breaking occurs for as few as two swimmers from each population (Fig.~\ref{figpar}e). The vortex and torque then increase approximately linearly with the number of swimmers $N_s$. 
For denser suspensions, the vortex destabilises the clusters into a mixed spinning core, and the torque $\tau$ plateaus at $\tau_p \approx \SI{0.6}{\nano\newton\micro\meter}$, consistently with experimental values $\tau \approx 0.2 - \SI{0.5}{\nano\newton\micro\meter}$ (Fig.~\ref{figpar}f).

Our simulations reproduce the experiments of Ref.~\cite{vincenti2019magnetotactic}, showing that hydrodynamic interactions are at the essence of the observed symmetry breaking.
Beyond this, our simulations allow us to explore a range of microscopic parameters to explain its underlying mechanisms.  
Notably, we study achiral bacteria (i.e.~no torque dipole, $\gamma=0$), and find that rotational symmetry is only broken if both NS and SS populations coexist. Even then, the system randomly develops either a CW or a CCW vortex, unlike the consistent CW pattern in experiments. 

In contrast, for moderately chiral bacteria ($\gamma>0$), symmetry breaking occurs CW systematically for either one (for ex, just NS) and two (NS and SS) populations. 
Remarkably, when either chirality is strong (large $\gamma$), or the force dipole $s$ and gravity $v_g$ are weak, the stationary state vanishes and gives way to oscillations. The clusters precess around the poles, and the flow alternates between CW and CCW. 

This suggests the presence of two distinct mechanisms causing the observed vortex: a global achiral one (G) in which NS and SS clusters interact at long range, and a local chiral one (L) that allows a single population to break symmetry.


\paragraph{(G) Global symmetry breaking.}
The first mechanism, achiral interactions between NS and SS clusters, involves only the leading-order force dipole signature of the swimmers. 
As shown in Figs.~\ref{fig1}d and \ref{figmec}a, extensile force dipoles ($s$) create a lateral attractive flow~\cite{Squires_2000}. 
 For aligned pusher swimmers parallel to one another, in particular for bacteria swimming perpendicular to a plane wall, this lateral flows results in a well-known attraction \cite{drescher2009dancing} and clustering \cite{Petroff_2015,Thery_2020}.  
As a result, for sufficiently strong fields the swimmers from a population in the drop align, attract, and form clusters that move cohesively on the boundary. 
For two populations, each cluster pushes flow away from its pole and thus destabilises the centre of mass of the cluster at the opposite pole. This configuration is unstable. Both clusters then move laterally, leading to broken symmetry and a vortical flow. 
Gravity acts as a stabiliser in the $z$-direction, so the instability occurs horizontally. Crucially, for achiral swimmers symmetry is broken randomly, and both CW and CCW vortices emerge with equal probability (Fig.~\ref{figmec}b).

\paragraph{(L) Local symmetry breaking.}
The systematic selection of CW rotation relies on a second physical mechanism. Our simulations reveal that it involves the higher-order effect of chirality $\gamma$ and occurs even for a single population (Fig.~\ref{figmec}c), hinting at the significance of local hydrodynamic interactions with the boundary. 
Under the assumption that the swimmers are aligned with the external field, a single population of, say NS, swimmers form a cluster, which moves cohesively along the droplet boundary. We analyse its trajectory by considering each contribution to the motion individually (Fig.~\ref{figmec}e). 
(i) Self-propulsion along the $+y$-direction drives swimmers towards the magnetic NP. (ii) Gravity uniformly displaces the cluster downwards to $-\hat{\mathbf{z}}$.  
(iii) Extensile force dipoles are hydrodynamically attracted to parallel surfaces through their image. In our geometry, this drives them to the $y=0$ plane, opposite to self-propulsion.
These three contributions are $\pm x$-symmetric and do not yet explain vortex formation. However, (iv) a torque dipole facing an inclined boundary creates a net lateral flow in its vicinity, oriented tangential to the boundary and orthogonal to the swimmer axis (Fig.~\ref{figmec}d). 
 While no individual swimmer can advect itself in this way, the collective action of bacteria on each other laterally displaces the cluster as a whole, while cohesion is maintained by swimmer-swimmer attraction. On the drop surface, this lateral translation results in a precession around the $y$-axis.

A steady-state position for the cluster then exists if these four effects balance locally, which can only occur at a shifted CW position (Fig.~\ref{figmec}f). If $\gamma$ (i.e.~ the contribution in (iv)) is too strong, there is no stationary point and the clusters precess around the poles, creating the oscillating regime we observe numerically.

Our simulations reveal therefore that the systematic CW vortex in experiments results from the combination of two distinct physical mechanisms: (G) a global achiral interaction between cell clusters leads to global broken symmetry with no preferred direction, and (L) a local chiral confinement-mediated mechanism leads to local CW symmetry breaking.


 \begin{figure}[t]
\centering
		\includegraphics[width = \columnwidth]{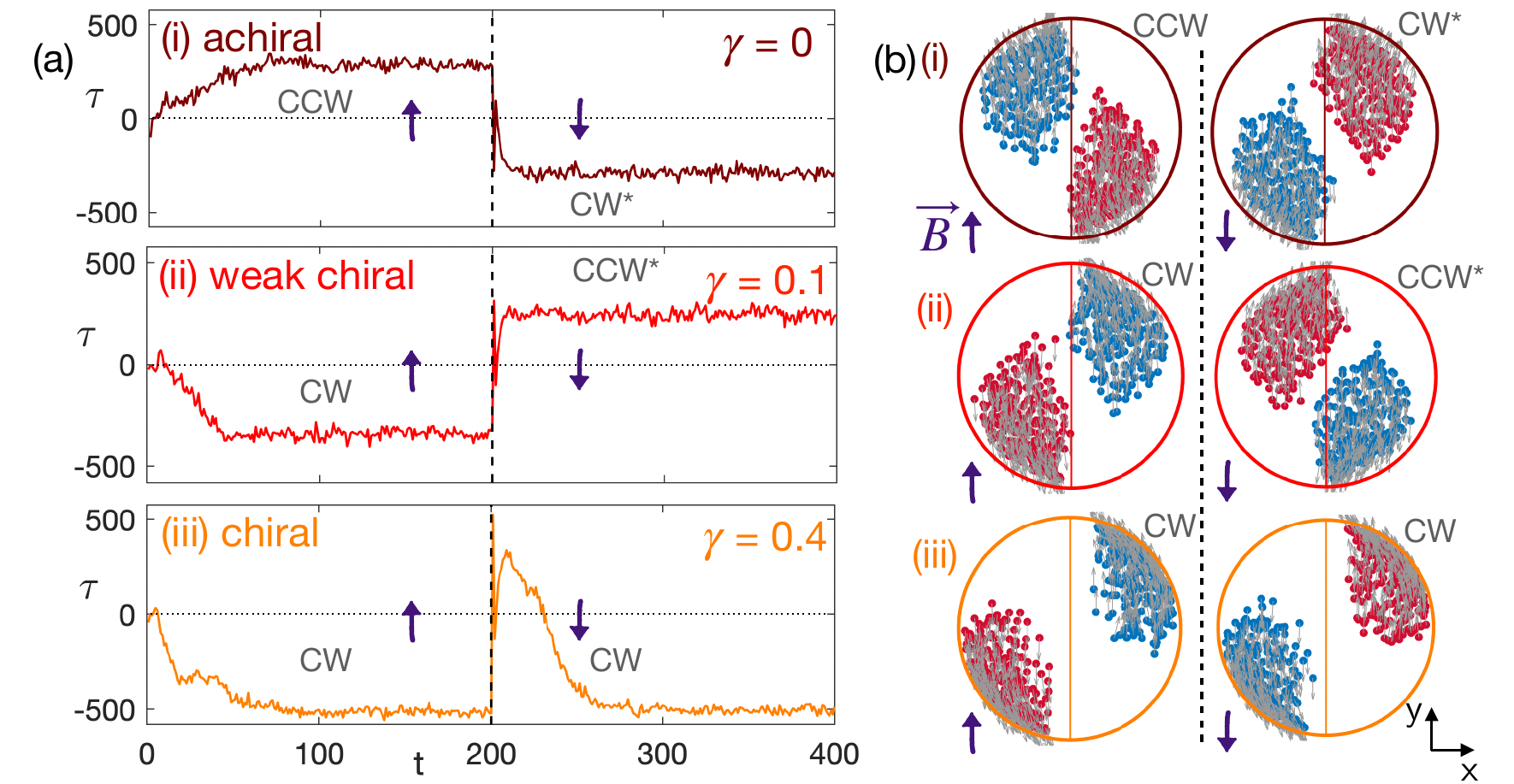}
		\caption{\label{fieldrev} Magnetic field reversal in (i) achiral, (ii) weakly chiral, (iii) chiral suspensions. (a) The reversal at $t = 200$ (dashed line) causes the reversal of the torque $\tau$ (denoted as $^*$) for achiral and weakly chiral swimmers (i, ii) while for a stronger chirality, the torque regains its initial CW direction (iii). 
		(b) Stationary-state position of swimmers before (left) and after (right) the field reversal. (i) The achiral system is symmetric with respect to the ($x-z$) plane. (ii) The weakly chiral has a preferred CW direction, which is lost after the reversal because of the (G) interaction between the populations. (iii) Stronger chirality (L) overcomes this repulsion and the populations shift back to CW. 
		}
\end{figure}

\paragraph{Field reversal.}
The coexistence of two instabilities explains the puzzling experimental observation that the vortex direction is reversed upon reversal of the magnetic field, although it initially rotates CW regardless of its orientation~\cite{vincenti2019magnetotactic}. 
In a uniform population, both instabilities cooperate in shifting the location of the clusters laterally, with the CW direction set by the local one.  
In contrast, when the field is reversed, it acts on  swimmers that are already asymmetrically distributed. The chiral CW instability (L) then competes with the population-population repulsion (G) that keeps clusters on the side they were already on, and the new steady state depends on the dominant mechanism. 
Below a threshold chirality $\gamma$ (which depends $B$ and $N_s$) the achiral population-population hydrodynamic repulsion (G) keeps swimmers on the same (left or right) side of the $y-z$ plane as they exchange poles: the flow is thus reversed to CCW, as in experiments (see Fig.~\ref{fieldrev}(i-ii)). 
Above this threshold, the chiral hydrodynamic effect (L) is sufficient for the populations to shift CW azimuthally and exchange sides, so the flow and torque eventually return to CW (Fig.~\ref{fieldrev}(iii)).

Reversing the field differentiates between swimmers where force propulsion dominates, such as MSR-1, and more chiral ones. Since increasing chirality further induces oscillations, this could be used as a proxy to estimate the relative importance of forces and torques in propulsion.
 This is particularly relevant because the degree of cellular chirality is difficult to access experimentally. To our knowledge, the dipole strengths for MSR-1 have not been measured.

Our fundamental model reveals how a sequence of symmetry-breaking controls the collective dynamics of confined biased swimmers and the emergence of a global vortex in the equatorial plane. Two distinct mechanisms can trigger collective motion, (G) globally through the long-range repulsion of NS and SS populations and (L) locally for chiral swimmers. In the latter, a few swimmers of a single population are enough to break the lateral symmetry because of hydrodynamic interactions with inclined walls. This tunable and externally controlled collective motion offers a promising framework for the analysis of populations of biased cells. 
The reversal dynamics and oscillations capture the underlying microscopic properties of the swimmers, in particular the relative strength of hydrodynamic forces and torques, which may otherwise be hard to measure. For physically relevant values, oscillations arise in the steady state of strongly chiral swimmers, while the vortex reversal upon field reversal seen experimentally in~\cite{vincenti2019magnetotactic} is identified as the signature of weak chirality. 
The rotation direction then depends on the history of the suspension even in a linear system.

\begin{acknowledgments}
This project has received funding from the European Research Council under the European Union's Horizon 2020 research and innovation program (Grant No.~682754 to E.L.). A.T. received support from the Simons Foundation through the Math + X grant awarded to the University of Pennsylvania.
\end{acknowledgments}

\bibliography{bibmtb.bib}

\providecommand{\noopsort}[1]{}\providecommand{\singleletter}[1]{#1}%
\begin{thebibliography}{42}%
\makeatletter
\providecommand \@ifxundefined [1]{%
 \@ifx{#1\undefined}
}%
\providecommand \@ifnum [1]{%
 \ifnum #1\expandafter \@firstoftwo
 \else \expandafter \@secondoftwo
 \fi
}%
\providecommand \@ifx [1]{%
 \ifx #1\expandafter \@firstoftwo
 \else \expandafter \@secondoftwo
 \fi
}%
\providecommand \natexlab [1]{#1}%
\providecommand \enquote  [1]{``#1''}%
\providecommand \bibnamefont  [1]{#1}%
\providecommand \bibfnamefont [1]{#1}%
\providecommand \citenamefont [1]{#1}%
\providecommand \href@noop [0]{\@secondoftwo}%
\providecommand \href [0]{\begingroup \@sanitize@url \@href}%
\providecommand \@href[1]{\@@startlink{#1}\@@href}%
\providecommand \@@href[1]{\endgroup#1\@@endlink}%
\providecommand \@sanitize@url [0]{\catcode `\\12\catcode `\$12\catcode `\&12\catcode `\#12\catcode `\^12\catcode `\_12\catcode `\%12\relax}%
\providecommand \@@startlink[1]{}%
\providecommand \@@endlink[0]{}%
\providecommand \url  [0]{\begingroup\@sanitize@url \@url }%
\providecommand \@url [1]{\endgroup\@href {#1}{\urlprefix }}%
\providecommand \urlprefix  [0]{URL }%
\providecommand \Eprint [0]{\href }%
\providecommand \doibase [0]{https://doi.org/}%
\providecommand \selectlanguage [0]{\@gobble}%
\providecommand \bibinfo  [0]{\@secondoftwo}%
\providecommand \bibfield  [0]{\@secondoftwo}%
\providecommand \translation [1]{[#1]}%
\providecommand \BibitemOpen [0]{}%
\providecommand \bibitemStop [0]{}%
\providecommand \bibitemNoStop [0]{.\EOS\space}%
\providecommand \EOS [0]{\spacefactor3000\relax}%
\providecommand \BibitemShut  [1]{\csname bibitem#1\endcsname}%
\let\auto@bib@innerbib\@empty
\bibitem [{\citenamefont {Bray}(2000)}]{braybook}%
  \BibitemOpen
  \bibfield  {author} {\bibinfo {author} {\bibfnamefont {D.}~\bibnamefont {Bray}},\ }\href@noop {} {\emph {\bibinfo {title} {Cell Movements}}}\ (\bibinfo  {publisher} {Garland Publishing},\ \bibinfo {address} {New York, NY},\ \bibinfo {year} {2000})\BibitemShut {NoStop}%
\bibitem [{\citenamefont {Soto}\ and\ \citenamefont {Golestanian}(2014)}]{soto2014self}%
  \BibitemOpen
  \bibfield  {author} {\bibinfo {author} {\bibfnamefont {R.}~\bibnamefont {Soto}}\ and\ \bibinfo {author} {\bibfnamefont {R.}~\bibnamefont {Golestanian}},\ }\bibfield  {title} {\bibinfo {title} {Self-assembly of catalytically active colloidal molecules: Tailoring activity through surface chemistry},\ }\href {https://doi.org/10.1103/PhysRevLett.112.068301} {\bibfield  {journal} {\bibinfo  {journal} {Phys. Rev. Lett.}\ }\textbf {\bibinfo {volume} {112}},\ \bibinfo {pages} {068301} (\bibinfo {year} {2014})}\BibitemShut {NoStop}%
\bibitem [{\citenamefont {Vincent}\ and\ \citenamefont {Hill}(1996)}]{vincent1996bioconvection}%
  \BibitemOpen
  \bibfield  {author} {\bibinfo {author} {\bibfnamefont {R.~V.}\ \bibnamefont {Vincent}}\ and\ \bibinfo {author} {\bibfnamefont {N.~A.}\ \bibnamefont {Hill}},\ }\bibfield  {title} {\bibinfo {title} {Bioconvection in a suspension of phototactic algae},\ }\href {https://doi.org/10.1017/s0022112096008579} {\bibfield  {journal} {\bibinfo  {journal} {Journal of Fluid Mechanics}\ }\textbf {\bibinfo {volume} {327}},\ \bibinfo {pages} {343} (\bibinfo {year} {1996})}\BibitemShut {NoStop}%
\bibitem [{\citenamefont {Liebchen}\ \emph {et~al.}(2018)\citenamefont {Liebchen}, \citenamefont {Monderkamp}, \citenamefont {ten Hagen},\ and\ \citenamefont {L\"owen}}]{Liebchen2018viscotaxis}%
  \BibitemOpen
  \bibfield  {author} {\bibinfo {author} {\bibfnamefont {B.}~\bibnamefont {Liebchen}}, \bibinfo {author} {\bibfnamefont {P.}~\bibnamefont {Monderkamp}}, \bibinfo {author} {\bibfnamefont {B.}~\bibnamefont {ten Hagen}},\ and\ \bibinfo {author} {\bibfnamefont {H.}~\bibnamefont {L\"owen}},\ }\bibfield  {title} {\bibinfo {title} {Viscotaxis: Microswimmer navigation in viscosity gradients},\ }\href {https://doi.org/10.1103/PhysRevLett.120.208002} {\bibfield  {journal} {\bibinfo  {journal} {Phys. Rev. Lett.}\ }\textbf {\bibinfo {volume} {120}},\ \bibinfo {pages} {208002} (\bibinfo {year} {2018})}\BibitemShut {NoStop}%
\bibitem [{\citenamefont {Pedley}(2015)}]{pedley_2015}%
  \BibitemOpen
  \bibfield  {author} {\bibinfo {author} {\bibfnamefont {T.~J.}\ \bibnamefont {Pedley}},\ }\bibfield  {title} {\bibinfo {title} {Gyrotaxis in uniform vorticity},\ }\href {https://doi.org/10.1017/jfm.2014.666} {\bibfield  {journal} {\bibinfo  {journal} {J. Fluid Mech.}\ }\textbf {\bibinfo {volume} {762}},\ \bibinfo {pages} {R6} (\bibinfo {year} {2015})}\BibitemShut {NoStop}%
\bibitem [{\citenamefont {Hill}\ and\ \citenamefont {Pedley}(2005{\natexlab{a}})}]{Hill_2005}%
  \BibitemOpen
  \bibfield  {author} {\bibinfo {author} {\bibfnamefont {N.~A.}\ \bibnamefont {Hill}}\ and\ \bibinfo {author} {\bibfnamefont {T.~J.}\ \bibnamefont {Pedley}},\ }\bibfield  {title} {\bibinfo {title} {Bioconvection},\ }\href {https://doi.org/10.1016/j.fluiddyn.2005.03.002} {\bibfield  {journal} {\bibinfo  {journal} {Fluid Dynamics Research}\ }\textbf {\bibinfo {volume} {37}},\ \bibinfo {pages} {1} (\bibinfo {year} {2005}{\natexlab{a}})}\BibitemShut {NoStop}%
\bibitem [{\citenamefont {Lauga}\ and\ \citenamefont {Nadal}(2016)}]{Lauga_2016}%
  \BibitemOpen
  \bibfield  {author} {\bibinfo {author} {\bibfnamefont {E.}~\bibnamefont {Lauga}}\ and\ \bibinfo {author} {\bibfnamefont {F.}~\bibnamefont {Nadal}},\ }\bibfield  {title} {\bibinfo {title} {Clustering instability of focused swimmers},\ }\href {https://doi.org/10.1209/0295-5075/116/64004} {\bibfield  {journal} {\bibinfo  {journal} {{EPL} (Europhysics Letters)}\ }\textbf {\bibinfo {volume} {116}},\ \bibinfo {pages} {64004} (\bibinfo {year} {2016})}\BibitemShut {NoStop}%
\bibitem [{\citenamefont {Blakemore}(1975)}]{Blakemore_1975}%
  \BibitemOpen
  \bibfield  {author} {\bibinfo {author} {\bibfnamefont {R.}~\bibnamefont {Blakemore}},\ }\bibfield  {title} {\bibinfo {title} {Magnetotactic bacteria},\ }\href {https://doi.org/10.1126/science.170679} {\bibfield  {journal} {\bibinfo  {journal} {Science}\ }\textbf {\bibinfo {volume} {190}},\ \bibinfo {pages} {377} (\bibinfo {year} {1975})}\BibitemShut {NoStop}%
\bibitem [{\citenamefont {Waisbord}\ \emph {et~al.}(2016)\citenamefont {Waisbord}, \citenamefont {Lef{\`e}vre}, \citenamefont {Bocquet}, \citenamefont {Ybert},\ and\ \citenamefont {Cottin-Bizonne}}]{waisbord2016destabilization}%
  \BibitemOpen
  \bibfield  {author} {\bibinfo {author} {\bibfnamefont {N.}~\bibnamefont {Waisbord}}, \bibinfo {author} {\bibfnamefont {C.~T.}\ \bibnamefont {Lef{\`e}vre}}, \bibinfo {author} {\bibfnamefont {L.}~\bibnamefont {Bocquet}}, \bibinfo {author} {\bibfnamefont {C.}~\bibnamefont {Ybert}},\ and\ \bibinfo {author} {\bibfnamefont {C.}~\bibnamefont {Cottin-Bizonne}},\ }\bibfield  {title} {\bibinfo {title} {Destabilization of a flow focused suspension of magnetotactic bacteria},\ }\href {https://doi.org/10.1103/physrevfluids.1.053203} {\bibfield  {journal} {\bibinfo  {journal} {Phys. Rev. Fluids}\ }\textbf {\bibinfo {volume} {1}},\ \bibinfo {pages} {053203} (\bibinfo {year} {2016})}\BibitemShut {NoStop}%
\bibitem [{\citenamefont {Meng}\ \emph {et~al.}(2018)\citenamefont {Meng}, \citenamefont {Matsunaga},\ and\ \citenamefont {Golestanian}}]{Meng2018}%
  \BibitemOpen
  \bibfield  {author} {\bibinfo {author} {\bibfnamefont {F.}~\bibnamefont {Meng}}, \bibinfo {author} {\bibfnamefont {D.}~\bibnamefont {Matsunaga}},\ and\ \bibinfo {author} {\bibfnamefont {R.}~\bibnamefont {Golestanian}},\ }\bibfield  {title} {\bibinfo {title} {Clustering of magnetic swimmers in a {Poiseuille} flow},\ }\href {https://doi.org/10.1103/PhysRevLett.120.188101} {\bibfield  {journal} {\bibinfo  {journal} {Phys. Rev. Lett.}\ }\textbf {\bibinfo {volume} {120}},\ \bibinfo {pages} {188101} (\bibinfo {year} {2018})}\BibitemShut {NoStop}%
\bibitem [{\citenamefont {Petroff}\ \emph {et~al.}(2015)\citenamefont {Petroff}, \citenamefont {Wu},\ and\ \citenamefont {Libchaber}}]{Petroff_2015}%
  \BibitemOpen
  \bibfield  {author} {\bibinfo {author} {\bibfnamefont {A.~P.}\ \bibnamefont {Petroff}}, \bibinfo {author} {\bibfnamefont {X.-L.}\ \bibnamefont {Wu}},\ and\ \bibinfo {author} {\bibfnamefont {A.}~\bibnamefont {Libchaber}},\ }\bibfield  {title} {\bibinfo {title} {Fast-moving bacteria self-organize into active two-dimensional crystals of rotating cells},\ }\bibfield  {journal} {\bibinfo  {journal} {Phys. Rev. Lett.}\ }\textbf {\bibinfo {volume} {114}},\ \href {https://doi.org/10.1103/physrevlett.114.158102} {10.1103/physrevlett.114.158102} (\bibinfo {year} {2015})\BibitemShut {NoStop}%
\bibitem [{\citenamefont {Pierce}\ \emph {et~al.}(2017)\citenamefont {Pierce}, \citenamefont {Mumper}, \citenamefont {Brown}, \citenamefont {Brangham}, \citenamefont {Lower}, \citenamefont {Lower}, \citenamefont {Yang},\ and\ \citenamefont {Sooryakumar}}]{pierce2017tuning}%
  \BibitemOpen
  \bibfield  {author} {\bibinfo {author} {\bibfnamefont {C.~J.}\ \bibnamefont {Pierce}}, \bibinfo {author} {\bibfnamefont {E.}~\bibnamefont {Mumper}}, \bibinfo {author} {\bibfnamefont {E.~E.}\ \bibnamefont {Brown}}, \bibinfo {author} {\bibfnamefont {J.~T.}\ \bibnamefont {Brangham}}, \bibinfo {author} {\bibfnamefont {B.~H.}\ \bibnamefont {Lower}}, \bibinfo {author} {\bibfnamefont {S.~K.}\ \bibnamefont {Lower}}, \bibinfo {author} {\bibfnamefont {F.~Y.}\ \bibnamefont {Yang}},\ and\ \bibinfo {author} {\bibfnamefont {R.}~\bibnamefont {Sooryakumar}},\ }\bibfield  {title} {\bibinfo {title} {Tuning bacterial hydrodynamics with magnetic fields},\ }\href {https://doi.org/10.1103/PhysRevE.95.062612} {\bibfield  {journal} {\bibinfo  {journal} {Phys. Rev. E}\ }\textbf {\bibinfo {volume} {95}},\ \bibinfo {pages} {062612} (\bibinfo {year} {2017})}\BibitemShut {NoStop}%
\bibitem [{\citenamefont {Pierce}\ \emph {et~al.}(2020)\citenamefont {Pierce}, \citenamefont {Wijesinghe}, \citenamefont {Osborne}, \citenamefont {Mumper}, \citenamefont {Lower}, \citenamefont {Lower},\ and\ \citenamefont {Sooryakumar}}]{Pierce_2020}%
  \BibitemOpen
  \bibfield  {author} {\bibinfo {author} {\bibfnamefont {C.~J.}\ \bibnamefont {Pierce}}, \bibinfo {author} {\bibfnamefont {H.}~\bibnamefont {Wijesinghe}}, \bibinfo {author} {\bibfnamefont {E.}~\bibnamefont {Osborne}}, \bibinfo {author} {\bibfnamefont {E.}~\bibnamefont {Mumper}}, \bibinfo {author} {\bibfnamefont {B.}~\bibnamefont {Lower}}, \bibinfo {author} {\bibfnamefont {S.}~\bibnamefont {Lower}},\ and\ \bibinfo {author} {\bibfnamefont {R.}~\bibnamefont {Sooryakumar}},\ }\bibfield  {title} {\bibinfo {title} {Tunable self-assembly of magnetotactic bacteria: Role of hydrodynamics and magnetism},\ }\href {https://doi.org/10.1063/1.5129925} {\bibfield  {journal} {\bibinfo  {journal} {{AIP} Adv.}\ }\textbf {\bibinfo {volume} {10}},\ \bibinfo {pages} {015335} (\bibinfo {year} {2020})}\BibitemShut {NoStop}%
\bibitem [{\citenamefont {Th{\'{e}}ry}\ \emph {et~al.}(2020)\citenamefont {Th{\'{e}}ry}, \citenamefont {Nagard}, \citenamefont {dit Biot}, \citenamefont {Fradin}, \citenamefont {Dalnoki-Veress},\ and\ \citenamefont {Lauga}}]{Thery_2020}%
  \BibitemOpen
  \bibfield  {author} {\bibinfo {author} {\bibfnamefont {A.}~\bibnamefont {Th{\'{e}}ry}}, \bibinfo {author} {\bibfnamefont {L.~L.}\ \bibnamefont {Nagard}}, \bibinfo {author} {\bibfnamefont {J.-C.~O.}\ \bibnamefont {dit Biot}}, \bibinfo {author} {\bibfnamefont {C.}~\bibnamefont {Fradin}}, \bibinfo {author} {\bibfnamefont {K.}~\bibnamefont {Dalnoki-Veress}},\ and\ \bibinfo {author} {\bibfnamefont {E.}~\bibnamefont {Lauga}},\ }\bibfield  {title} {\bibinfo {title} {Self-organisation and convection of confined magnetotactic bacteria},\ }\bibfield  {journal} {\bibinfo  {journal} {Sci. Rep.}\ }\textbf {\bibinfo {volume} {10}},\ \href {https://doi.org/10.1038/s41598-020-70270-0} {10.1038/s41598-020-70270-0} (\bibinfo {year} {2020})\BibitemShut {NoStop}%
\bibitem [{\citenamefont {Pedley}\ and\ \citenamefont {Kessler}(1992)}]{pedley1992hydrodynamic}%
  \BibitemOpen
  \bibfield  {author} {\bibinfo {author} {\bibfnamefont {T.}~\bibnamefont {Pedley}}\ and\ \bibinfo {author} {\bibfnamefont {J.}~\bibnamefont {Kessler}},\ }\bibfield  {title} {\bibinfo {title} {Hydrodynamic phenomena in suspensions of swimming microorganisms},\ }\href {https://doi.org/10.1146/annurev.fl.24.010192.001525} {\bibfield  {journal} {\bibinfo  {journal} {Annu. Rev. Fluid Mech.}\ }\textbf {\bibinfo {volume} {24}},\ \bibinfo {pages} {313} (\bibinfo {year} {1992})}\BibitemShut {NoStop}%
\bibitem [{\citenamefont {Hill}\ and\ \citenamefont {Pedley}(2005{\natexlab{b}})}]{hill2005bioconvection}%
  \BibitemOpen
  \bibfield  {author} {\bibinfo {author} {\bibfnamefont {N.}~\bibnamefont {Hill}}\ and\ \bibinfo {author} {\bibfnamefont {T.}~\bibnamefont {Pedley}},\ }\bibfield  {title} {\bibinfo {title} {Bioconvection},\ }\href {https://doi.org/10.1016/j.fluiddyn.2005.03.002} {\bibfield  {journal} {\bibinfo  {journal} {Fluid Dyn. Res.}\ }\textbf {\bibinfo {volume} {37}},\ \bibinfo {pages} {1} (\bibinfo {year} {2005}{\natexlab{b}})}\BibitemShut {NoStop}%
\bibitem [{\citenamefont {Vincenti}\ \emph {et~al.}(2019)\citenamefont {Vincenti}, \citenamefont {Ramos}, \citenamefont {Cordero}, \citenamefont {Douarche}, \citenamefont {Soto},\ and\ \citenamefont {Cl{\'e}ment}}]{vincenti2019magnetotactic}%
  \BibitemOpen
  \bibfield  {author} {\bibinfo {author} {\bibfnamefont {B.}~\bibnamefont {Vincenti}}, \bibinfo {author} {\bibfnamefont {G.}~\bibnamefont {Ramos}}, \bibinfo {author} {\bibfnamefont {M.~L.}\ \bibnamefont {Cordero}}, \bibinfo {author} {\bibfnamefont {C.}~\bibnamefont {Douarche}}, \bibinfo {author} {\bibfnamefont {R.}~\bibnamefont {Soto}},\ and\ \bibinfo {author} {\bibfnamefont {E.}~\bibnamefont {Cl{\'e}ment}},\ }\bibfield  {title} {\bibinfo {title} {Magnetotactic bacteria in a droplet self-assemble into a rotary motor},\ }\href {https://doi.org/10.1038/s41467-019-13031-6} {\bibfield  {journal} {\bibinfo  {journal} {Nat. Commun.}\ }\textbf {\bibinfo {volume} {10}},\ \bibinfo {pages} {1} (\bibinfo {year} {2019})}\BibitemShut {NoStop}%
\bibitem [{\citenamefont {Opathalage}\ \emph {et~al.}(2019)\citenamefont {Opathalage}, \citenamefont {Norton}, \citenamefont {Juniper}, \citenamefont {Langeslay}, \citenamefont {Aghvami}, \citenamefont {Fraden},\ and\ \citenamefont {Dogic}}]{Opathalage_2019}%
  \BibitemOpen
  \bibfield  {author} {\bibinfo {author} {\bibfnamefont {A.}~\bibnamefont {Opathalage}}, \bibinfo {author} {\bibfnamefont {M.~M.}\ \bibnamefont {Norton}}, \bibinfo {author} {\bibfnamefont {M.~P.~N.}\ \bibnamefont {Juniper}}, \bibinfo {author} {\bibfnamefont {B.}~\bibnamefont {Langeslay}}, \bibinfo {author} {\bibfnamefont {S.~A.}\ \bibnamefont {Aghvami}}, \bibinfo {author} {\bibfnamefont {S.}~\bibnamefont {Fraden}},\ and\ \bibinfo {author} {\bibfnamefont {Z.}~\bibnamefont {Dogic}},\ }\bibfield  {title} {\bibinfo {title} {Self-organized dynamics and the transition to turbulence of confined active nematics},\ }\href {https://doi.org/10.1073/pnas.1816733116} {\bibfield  {journal} {\bibinfo  {journal} {Proceedings of the National Academy of Sciences}\ }\textbf {\bibinfo {volume} {116}},\ \bibinfo {pages} {4788} (\bibinfo {year} {2019})}\BibitemShut {NoStop}%
\bibitem [{\citenamefont {Chen}\ \emph {et~al.}(2021)\citenamefont {Chen}, \citenamefont {Mani}, \citenamefont {Karani}, \citenamefont {Li}, \citenamefont {Mani},\ and\ \citenamefont {Tang}}]{Chen_2021}%
  \BibitemOpen
  \bibfield  {author} {\bibinfo {author} {\bibfnamefont {W.}~\bibnamefont {Chen}}, \bibinfo {author} {\bibfnamefont {N.}~\bibnamefont {Mani}}, \bibinfo {author} {\bibfnamefont {H.}~\bibnamefont {Karani}}, \bibinfo {author} {\bibfnamefont {H.}~\bibnamefont {Li}}, \bibinfo {author} {\bibfnamefont {S.}~\bibnamefont {Mani}},\ and\ \bibinfo {author} {\bibfnamefont {J.~X.}\ \bibnamefont {Tang}},\ }\bibfield  {title} {\bibinfo {title} {Confinement discerns swarmers from planktonic bacteria},\ }\bibfield  {journal} {\bibinfo  {journal} {{eLife}}\ }\textbf {\bibinfo {volume} {10}},\ \href {https://doi.org/10.7554/elife.64176} {10.7554/elife.64176} (\bibinfo {year} {2021})\BibitemShut {NoStop}%
\bibitem [{\citenamefont {Huang}\ \emph {et~al.}(2021)\citenamefont {Huang}, \citenamefont {Du}, \citenamefont {Jiang},\ and\ \citenamefont {Hou}}]{Huang_2021}%
  \BibitemOpen
  \bibfield  {author} {\bibinfo {author} {\bibfnamefont {D.}~\bibnamefont {Huang}}, \bibinfo {author} {\bibfnamefont {Y.}~\bibnamefont {Du}}, \bibinfo {author} {\bibfnamefont {H.}~\bibnamefont {Jiang}},\ and\ \bibinfo {author} {\bibfnamefont {Z.}~\bibnamefont {Hou}},\ }\bibfield  {title} {\bibinfo {title} {Emergent spiral vortex of confined biased active particles},\ }\href {https://doi.org/10.1103/PhysRevE.104.034606} {\bibfield  {journal} {\bibinfo  {journal} {Phys. Rev. E}\ }\textbf {\bibinfo {volume} {104}},\ \bibinfo {pages} {034606} (\bibinfo {year} {2021})}\BibitemShut {NoStop}%
\bibitem [{\citenamefont {Lushi}\ \emph {et~al.}(2014)\citenamefont {Lushi}, \citenamefont {Wioland},\ and\ \citenamefont {Goldstein}}]{lushi2014fluid}%
  \BibitemOpen
  \bibfield  {author} {\bibinfo {author} {\bibfnamefont {E.}~\bibnamefont {Lushi}}, \bibinfo {author} {\bibfnamefont {H.}~\bibnamefont {Wioland}},\ and\ \bibinfo {author} {\bibfnamefont {R.~E.}\ \bibnamefont {Goldstein}},\ }\bibfield  {title} {\bibinfo {title} {Fluid flows created by swimming bacteria drive self-organization in confined suspensions},\ }\href {https://doi.org/10.1073/pnas.1405698111} {\bibfield  {journal} {\bibinfo  {journal} {Proc. Natl. Acad. Sci. U. S. A}\ ,\ \bibinfo {pages} {201405698}} (\bibinfo {year} {2014})}\BibitemShut {NoStop}%
\bibitem [{\citenamefont {Wioland}\ \emph {et~al.}(2013)\citenamefont {Wioland}, \citenamefont {Woodhouse}, \citenamefont {Dunkel}, \citenamefont {Kessler},\ and\ \citenamefont {Goldstein}}]{wioland2013confinement}%
  \BibitemOpen
  \bibfield  {author} {\bibinfo {author} {\bibfnamefont {H.}~\bibnamefont {Wioland}}, \bibinfo {author} {\bibfnamefont {F.~G.}\ \bibnamefont {Woodhouse}}, \bibinfo {author} {\bibfnamefont {J.}~\bibnamefont {Dunkel}}, \bibinfo {author} {\bibfnamefont {J.~O.}\ \bibnamefont {Kessler}},\ and\ \bibinfo {author} {\bibfnamefont {R.~E.}\ \bibnamefont {Goldstein}},\ }\bibfield  {title} {\bibinfo {title} {Confinement stabilizes a bacterial suspension into a spiral vortex},\ }\href@noop {} {\bibfield  {journal} {\bibinfo  {journal} {Phys. Rev. Lett.}\ }\textbf {\bibinfo {volume} {110}},\ \bibinfo {pages} {268102} (\bibinfo {year} {2013})}\BibitemShut {NoStop}%
\bibitem [{\citenamefont {Vladescu}\ \emph {et~al.}(2014)\citenamefont {Vladescu}, \citenamefont {Marsden}, \citenamefont {Schwarz-Linek}, \citenamefont {Martinez}, \citenamefont {Arlt}, \citenamefont {Morozov}, \citenamefont {Marenduzzo}, \citenamefont {Cates},\ and\ \citenamefont {Poon}}]{Vladescu2014}%
  \BibitemOpen
  \bibfield  {author} {\bibinfo {author} {\bibfnamefont {I.~D.}\ \bibnamefont {Vladescu}}, \bibinfo {author} {\bibfnamefont {E.~J.}\ \bibnamefont {Marsden}}, \bibinfo {author} {\bibfnamefont {J.}~\bibnamefont {Schwarz-Linek}}, \bibinfo {author} {\bibfnamefont {V.~A.}\ \bibnamefont {Martinez}}, \bibinfo {author} {\bibfnamefont {J.}~\bibnamefont {Arlt}}, \bibinfo {author} {\bibfnamefont {A.~N.}\ \bibnamefont {Morozov}}, \bibinfo {author} {\bibfnamefont {D.}~\bibnamefont {Marenduzzo}}, \bibinfo {author} {\bibfnamefont {M.~E.}\ \bibnamefont {Cates}},\ and\ \bibinfo {author} {\bibfnamefont {W.~C.~K.}\ \bibnamefont {Poon}},\ }\bibfield  {title} {\bibinfo {title} {Filling an emulsion drop with motile bacteria},\ }\href {https://doi.org/10.1103/PhysRevLett.113.268101} {\bibfield  {journal} {\bibinfo  {journal} {Phys. Rev. Lett.}\ }\textbf {\bibinfo {volume} {113}},\ \bibinfo {pages} {268101} (\bibinfo {year} {2014})}\BibitemShut {NoStop}%
\bibitem [{\citenamefont {Hamby}\ \emph {et~al.}(2018)\citenamefont {Hamby}, \citenamefont {Vig}, \citenamefont {Safonova},\ and\ \citenamefont {Wolgemuth}}]{Hamby_2018}%
  \BibitemOpen
  \bibfield  {author} {\bibinfo {author} {\bibfnamefont {A.~E.}\ \bibnamefont {Hamby}}, \bibinfo {author} {\bibfnamefont {D.~K.}\ \bibnamefont {Vig}}, \bibinfo {author} {\bibfnamefont {S.}~\bibnamefont {Safonova}},\ and\ \bibinfo {author} {\bibfnamefont {C.~W.}\ \bibnamefont {Wolgemuth}},\ }\bibfield  {title} {\bibinfo {title} {Swimming bacteria power microspin cycles},\ }\bibfield  {journal} {\bibinfo  {journal} {Sci. Adv.}\ }\textbf {\bibinfo {volume} {4}},\ \href {https://doi.org/10.1126/sciadv.aau0125} {10.1126/sciadv.aau0125} (\bibinfo {year} {2018})\BibitemShut {NoStop}%
\bibitem [{\citenamefont {Creppy}\ \emph {et~al.}(2016)\citenamefont {Creppy}, \citenamefont {Plourabou{\'{e}}}, \citenamefont {Praud}, \citenamefont {Druart}, \citenamefont {Cazin}, \citenamefont {Yu},\ and\ \citenamefont {Degond}}]{Creppy_2016}%
  \BibitemOpen
  \bibfield  {author} {\bibinfo {author} {\bibfnamefont {A.}~\bibnamefont {Creppy}}, \bibinfo {author} {\bibfnamefont {F.}~\bibnamefont {Plourabou{\'{e}}}}, \bibinfo {author} {\bibfnamefont {O.}~\bibnamefont {Praud}}, \bibinfo {author} {\bibfnamefont {X.}~\bibnamefont {Druart}}, \bibinfo {author} {\bibfnamefont {S.}~\bibnamefont {Cazin}}, \bibinfo {author} {\bibfnamefont {H.}~\bibnamefont {Yu}},\ and\ \bibinfo {author} {\bibfnamefont {P.}~\bibnamefont {Degond}},\ }\bibfield  {title} {\bibinfo {title} {Symmetry-breaking phase transitions in highly concentrated semen},\ }\href {https://doi.org/10.1098/rsif.2016.0575} {\bibfield  {journal} {\bibinfo  {journal} {J. R. Soc. Interface}\ }\textbf {\bibinfo {volume} {13}},\ \bibinfo {pages} {20160575} (\bibinfo {year} {2016})}\BibitemShut {NoStop}%
\bibitem [{\citenamefont {Tsang}\ and\ \citenamefont {Kanso}(2015)}]{Tsang_2015}%
  \BibitemOpen
  \bibfield  {author} {\bibinfo {author} {\bibfnamefont {A.~C.~H.}\ \bibnamefont {Tsang}}\ and\ \bibinfo {author} {\bibfnamefont {E.}~\bibnamefont {Kanso}},\ }\bibfield  {title} {\bibinfo {title} {Circularly confined microswimmers exhibit multiple global patterns},\ }\bibfield  {journal} {\bibinfo  {journal} {Phys. Rev. E}\ }\textbf {\bibinfo {volume} {91}},\ \href {https://doi.org/10.1103/physreve.91.043008} {10.1103/physreve.91.043008} (\bibinfo {year} {2015})\BibitemShut {NoStop}%
\bibitem [{\citenamefont {Lauga}\ and\ \citenamefont {Powers}(2009)}]{lauga2009hydrodynamics}%
  \BibitemOpen
  \bibfield  {author} {\bibinfo {author} {\bibfnamefont {E.}~\bibnamefont {Lauga}}\ and\ \bibinfo {author} {\bibfnamefont {T.~R.}\ \bibnamefont {Powers}},\ }\bibfield  {title} {\bibinfo {title} {The hydrodynamics of swimming microorganisms},\ }\href {https://doi.org/10.1088/0034-4885/72/9/096601} {\bibfield  {journal} {\bibinfo  {journal} {Rep. Prog. Phys.}\ }\textbf {\bibinfo {volume} {72}},\ \bibinfo {pages} {096601} (\bibinfo {year} {2009})}\BibitemShut {NoStop}%
\bibitem [{\citenamefont {Reufer}\ \emph {et~al.}(2014)\citenamefont {Reufer}, \citenamefont {Besseling}, \citenamefont {Schwarz-Linek}, \citenamefont {Martinez}, \citenamefont {Morozov}, \citenamefont {Arlt}, \citenamefont {Trubitsyn}, \citenamefont {Ward},\ and\ \citenamefont {Poon}}]{Reufer_2014}%
  \BibitemOpen
  \bibfield  {author} {\bibinfo {author} {\bibfnamefont {M.}~\bibnamefont {Reufer}}, \bibinfo {author} {\bibfnamefont {R.}~\bibnamefont {Besseling}}, \bibinfo {author} {\bibfnamefont {J.}~\bibnamefont {Schwarz-Linek}}, \bibinfo {author} {\bibfnamefont {V.}~\bibnamefont {Martinez}}, \bibinfo {author} {\bibfnamefont {A.}~\bibnamefont {Morozov}}, \bibinfo {author} {\bibfnamefont {J.}~\bibnamefont {Arlt}}, \bibinfo {author} {\bibfnamefont {D.}~\bibnamefont {Trubitsyn}}, \bibinfo {author} {\bibfnamefont {F.}~\bibnamefont {Ward}},\ and\ \bibinfo {author} {\bibfnamefont {W.}~\bibnamefont {Poon}},\ }\bibfield  {title} {\bibinfo {title} {Switching of swimming modes in \textit{{Magnetospirillium} gryphiswaldense}},\ }\href {https://doi.org/10.1016/j.bpj.2013.10.038} {\bibfield  {journal} {\bibinfo  {journal} {Biophys. J.}\ }\textbf {\bibinfo {volume} {106}},\ \bibinfo {pages} {37} (\bibinfo {year} {2014})}\BibitemShut {NoStop}%
\bibitem [{\citenamefont {Maul}\ and\ \citenamefont {Kim}(1994)}]{maul1994image}%
  \BibitemOpen
  \bibfield  {author} {\bibinfo {author} {\bibfnamefont {C.}~\bibnamefont {Maul}}\ and\ \bibinfo {author} {\bibfnamefont {S.}~\bibnamefont {Kim}},\ }\bibfield  {title} {\bibinfo {title} {Image systems for a {Stokeslet} inside a rigid spherical container},\ }\href {https://doi.org/10.1063/1.868223} {\bibfield  {journal} {\bibinfo  {journal} {Phys. Fluids}\ }\textbf {\bibinfo {volume} {6}},\ \bibinfo {pages} {2221} (\bibinfo {year} {1994})}\BibitemShut {NoStop}%
\bibitem [{\citenamefont {Shail}(1987)}]{shail1987note}%
  \BibitemOpen
  \bibfield  {author} {\bibinfo {author} {\bibfnamefont {R.}~\bibnamefont {Shail}},\ }\bibfield  {title} {\bibinfo {title} {A note on some asymmetric {Stokes} flows within a sphere},\ }\href {https://doi.org/10.1093/qjmam/40.2.223} {\bibfield  {journal} {\bibinfo  {journal} {Q. J. Mech. Appl. Math.}\ }\textbf {\bibinfo {volume} {40}},\ \bibinfo {pages} {223} (\bibinfo {year} {1987})}\BibitemShut {NoStop}%
\bibitem [{\citenamefont {Hackborn}\ \emph {et~al.}(1986)\citenamefont {Hackborn}, \citenamefont {O'Neill},\ and\ \citenamefont {Ranger}}]{hackborn1986structure}%
  \BibitemOpen
  \bibfield  {author} {\bibinfo {author} {\bibfnamefont {W.}~\bibnamefont {Hackborn}}, \bibinfo {author} {\bibfnamefont {M.}~\bibnamefont {O'Neill}},\ and\ \bibinfo {author} {\bibfnamefont {K.}~\bibnamefont {Ranger}},\ }\bibfield  {title} {\bibinfo {title} {The structure of an asymmetric {Stokes} flow},\ }\href {https://doi.org/0.1093/qjmam/39.1.1} {\bibfield  {journal} {\bibinfo  {journal} {Q. J. Mech. Appl. Math.}\ }\textbf {\bibinfo {volume} {39}},\ \bibinfo {pages} {1} (\bibinfo {year} {1986})}\BibitemShut {NoStop}%
\bibitem [{\citenamefont {Chamolly}\ and\ \citenamefont {Lauga}(2020)}]{chamolly2020stokes}%
  \BibitemOpen
  \bibfield  {author} {\bibinfo {author} {\bibfnamefont {A.}~\bibnamefont {Chamolly}}\ and\ \bibinfo {author} {\bibfnamefont {E.}~\bibnamefont {Lauga}},\ }\bibfield  {title} {\bibinfo {title} {Stokes flow due to point torques and sources in a spherical geometry},\ }\href {https://doi.org/10.1103/PhysRevFluids.5.074202} {\bibfield  {journal} {\bibinfo  {journal} {Phys. Rev. Fluids}\ }\textbf {\bibinfo {volume} {5}},\ \bibinfo {pages} {074202} (\bibinfo {year} {2020})}\BibitemShut {NoStop}%
\bibitem [{\citenamefont {Zahn}\ \emph {et~al.}(2017)\citenamefont {Zahn}, \citenamefont {Keller}, \citenamefont {Toro-Nahuelpan}, \citenamefont {Dorscht}, \citenamefont {Gross}, \citenamefont {Laumann}, \citenamefont {Gekle}, \citenamefont {Zimmermann}, \citenamefont {Sch{\"u}ler},\ and\ \citenamefont {Kress}}]{Zahn_2017}%
  \BibitemOpen
  \bibfield  {author} {\bibinfo {author} {\bibfnamefont {C.}~\bibnamefont {Zahn}}, \bibinfo {author} {\bibfnamefont {S.}~\bibnamefont {Keller}}, \bibinfo {author} {\bibfnamefont {M.}~\bibnamefont {Toro-Nahuelpan}}, \bibinfo {author} {\bibfnamefont {P.}~\bibnamefont {Dorscht}}, \bibinfo {author} {\bibfnamefont {W.}~\bibnamefont {Gross}}, \bibinfo {author} {\bibfnamefont {M.}~\bibnamefont {Laumann}}, \bibinfo {author} {\bibfnamefont {S.}~\bibnamefont {Gekle}}, \bibinfo {author} {\bibfnamefont {W.}~\bibnamefont {Zimmermann}}, \bibinfo {author} {\bibfnamefont {D.}~\bibnamefont {Sch{\"u}ler}},\ and\ \bibinfo {author} {\bibfnamefont {H.}~\bibnamefont {Kress}},\ }\bibfield  {title} {\bibinfo {title} {Measurement of the magnetic moment of single \textit{{Magnetospirillium} gryphiswaldense} cells by magnetic tweezers},\ }\bibfield  {journal} {\bibinfo  {journal} {Sci. Rep.}\ }\textbf {\bibinfo {volume} {7}},\ \href {https://doi.org/10.1038/s41598-017-03756-z} {10.1038/s41598-017-03756-z} (\bibinfo {year}
  {2017})\BibitemShut {NoStop}%
\bibitem [{\citenamefont {Waisbord}\ \emph {et~al.}(2021)\citenamefont {Waisbord}, \citenamefont {Dehkharghani},\ and\ \citenamefont {Guasto}}]{Waisbord_2021}%
  \BibitemOpen
  \bibfield  {author} {\bibinfo {author} {\bibfnamefont {N.}~\bibnamefont {Waisbord}}, \bibinfo {author} {\bibfnamefont {A.}~\bibnamefont {Dehkharghani}},\ and\ \bibinfo {author} {\bibfnamefont {J.~S.}\ \bibnamefont {Guasto}},\ }\bibfield  {title} {\bibinfo {title} {Fluidic bacterial diodes rectify magnetotactic cell motility in porous environments},\ }\bibfield  {journal} {\bibinfo  {journal} {Nat. Commun.}\ }\textbf {\bibinfo {volume} {12}},\ \href {https://doi.org/10.1038/s41467-021-26235-6} {10.1038/s41467-021-26235-6} (\bibinfo {year} {2021})\BibitemShut {NoStop}%
\bibitem [{\citenamefont {Ford}\ and\ \citenamefont {Harvey}(2007)}]{Ford_2007}%
  \BibitemOpen
  \bibfield  {author} {\bibinfo {author} {\bibfnamefont {R.~M.}\ \bibnamefont {Ford}}\ and\ \bibinfo {author} {\bibfnamefont {R.~W.}\ \bibnamefont {Harvey}},\ }\bibfield  {title} {\bibinfo {title} {Role of chemotaxis in the transport of bacteria through saturated porous media},\ }\href {https://doi.org/10.1016/j.advwatres.2006.05.019} {\bibfield  {journal} {\bibinfo  {journal} {Adv. Water Resour.}\ }\textbf {\bibinfo {volume} {30}},\ \bibinfo {pages} {1608} (\bibinfo {year} {2007})}\BibitemShut {NoStop}%
\bibitem [{\citenamefont {Hu}\ \emph {et~al.}(2015)\citenamefont {Hu}, \citenamefont {Yang}, \citenamefont {Gompper},\ and\ \citenamefont {Winkler}}]{Hu_2015}%
  \BibitemOpen
  \bibfield  {author} {\bibinfo {author} {\bibfnamefont {J.}~\bibnamefont {Hu}}, \bibinfo {author} {\bibfnamefont {M.}~\bibnamefont {Yang}}, \bibinfo {author} {\bibfnamefont {G.}~\bibnamefont {Gompper}},\ and\ \bibinfo {author} {\bibfnamefont {R.~G.}\ \bibnamefont {Winkler}},\ }\bibfield  {title} {\bibinfo {title} {Modelling the mechanics and hydrodynamics of swimming \textit{{E}. coli}},\ }\href {https://doi.org/10.1039/c5sm01678a} {\bibfield  {journal} {\bibinfo  {journal} {Soft Matter}\ }\textbf {\bibinfo {volume} {11}},\ \bibinfo {pages} {7867} (\bibinfo {year} {2015})}\BibitemShut {NoStop}%
\bibitem [{\citenamefont {Squires}\ and\ \citenamefont {Brenner}(2000)}]{Squires_2000}%
  \BibitemOpen
  \bibfield  {author} {\bibinfo {author} {\bibfnamefont {T.~M.}\ \bibnamefont {Squires}}\ and\ \bibinfo {author} {\bibfnamefont {M.~P.}\ \bibnamefont {Brenner}},\ }\bibfield  {title} {\bibinfo {title} {Like-charge attraction and hydrodynamic interaction},\ }\href {https://doi.org/10.1103/physrevlett.85.4976} {\bibfield  {journal} {\bibinfo  {journal} {Phys. Rev. Lett.}\ }\textbf {\bibinfo {volume} {85}},\ \bibinfo {pages} {4976} (\bibinfo {year} {2000})}\BibitemShut {NoStop}%
\bibitem [{\citenamefont {Drescher}\ \emph {et~al.}(2009)\citenamefont {Drescher}, \citenamefont {Leptos}, \citenamefont {Tuval}, \citenamefont {Ishikawa}, \citenamefont {Pedley},\ and\ \citenamefont {Goldstein}}]{drescher2009dancing}%
  \BibitemOpen
  \bibfield  {author} {\bibinfo {author} {\bibfnamefont {K.}~\bibnamefont {Drescher}}, \bibinfo {author} {\bibfnamefont {K.~C.}\ \bibnamefont {Leptos}}, \bibinfo {author} {\bibfnamefont {I.}~\bibnamefont {Tuval}}, \bibinfo {author} {\bibfnamefont {T.}~\bibnamefont {Ishikawa}}, \bibinfo {author} {\bibfnamefont {T.~J.}\ \bibnamefont {Pedley}},\ and\ \bibinfo {author} {\bibfnamefont {R.~E.}\ \bibnamefont {Goldstein}},\ }\bibfield  {title} {\bibinfo {title} {Dancing \textit{{Volvox}}: hydrodynamic bound states of swimming algae},\ }\href {https://doi.org/10.1103/physrevlett.102.168101} {\bibfield  {journal} {\bibinfo  {journal} {Phys. Rev. Lett.}\ }\textbf {\bibinfo {volume} {102}},\ \bibinfo {pages} {168101} (\bibinfo {year} {2009})}\BibitemShut {NoStop}%
\bibitem [{\citenamefont {Muller}(1959)}]{muller1959note}%
  \BibitemOpen
  \bibfield  {author} {\bibinfo {author} {\bibfnamefont {M.~E.}\ \bibnamefont {Muller}},\ }\bibfield  {title} {\bibinfo {title} {A note on a method for generating points uniformly on {N}-dimensional spheres},\ }\href {https://doi.org/10.1145/377939.377946} {\bibfield  {journal} {\bibinfo  {journal} {Commun. ACM}\ }\textbf {\bibinfo {volume} {2}},\ \bibinfo {pages} {19} (\bibinfo {year} {1959})}\BibitemShut {NoStop}%
\bibitem [{\citenamefont {Drescher}\ \emph {et~al.}(2011)\citenamefont {Drescher}, \citenamefont {Dunkel}, \citenamefont {Cisneros}, \citenamefont {Ganguly},\ and\ \citenamefont {Goldstein}}]{drescher2011fluid}%
  \BibitemOpen
  \bibfield  {author} {\bibinfo {author} {\bibfnamefont {K.}~\bibnamefont {Drescher}}, \bibinfo {author} {\bibfnamefont {J.}~\bibnamefont {Dunkel}}, \bibinfo {author} {\bibfnamefont {L.~H.}\ \bibnamefont {Cisneros}}, \bibinfo {author} {\bibfnamefont {S.}~\bibnamefont {Ganguly}},\ and\ \bibinfo {author} {\bibfnamefont {R.~E.}\ \bibnamefont {Goldstein}},\ }\bibfield  {title} {\bibinfo {title} {Fluid dynamics and noise in bacterial cell--cell and cell--surface scattering},\ }\href {https://doi.org/10.1073/pnas.1019079108} {\bibfield  {journal} {\bibinfo  {journal} {Proc. Natl. Acad. Sci. U. S. A}\ }\textbf {\bibinfo {volume} {108}},\ \bibinfo {pages} {10940} (\bibinfo {year} {2011})}\BibitemShut {NoStop}%
\bibitem [{\citenamefont {Guazzelli}\ and\ \citenamefont {Morris}(2011)}]{guazzelli2011physical}%
  \BibitemOpen
  \bibfield  {author} {\bibinfo {author} {\bibfnamefont {E.}~\bibnamefont {Guazzelli}}\ and\ \bibinfo {author} {\bibfnamefont {J.~F.}\ \bibnamefont {Morris}},\ }\href {https://doi.org/10.1017/CBO9780511894671} {\emph {\bibinfo {title} {A physical introduction to suspension dynamics}}},\ Vol.~\bibinfo {volume} {45}\ (\bibinfo  {publisher} {Cambridge University Press},\ \bibinfo {year} {2011})\BibitemShut {NoStop}%
\bibitem [{\citenamefont {Z{\"o}ttl}\ \emph {et~al.}(2019)\citenamefont {Z{\"o}ttl}, \citenamefont {Klop}, \citenamefont {Balin}, \citenamefont {Gao}, \citenamefont {Yeomans},\ and\ \citenamefont {Aarts}}]{Zottl_2019}%
  \BibitemOpen
  \bibfield  {author} {\bibinfo {author} {\bibfnamefont {A.}~\bibnamefont {Z{\"o}ttl}}, \bibinfo {author} {\bibfnamefont {K.~E.}\ \bibnamefont {Klop}}, \bibinfo {author} {\bibfnamefont {A.~K.}\ \bibnamefont {Balin}}, \bibinfo {author} {\bibfnamefont {Y.}~\bibnamefont {Gao}}, \bibinfo {author} {\bibfnamefont {J.~M.}\ \bibnamefont {Yeomans}},\ and\ \bibinfo {author} {\bibfnamefont {D.~G. A.~L.}\ \bibnamefont {Aarts}},\ }\bibfield  {title} {\bibinfo {title} {Dynamics of individual {Brownian} rods in a microchannel flow},\ }\href {https://doi.org/10.1039/c9sm00903e} {\bibfield  {journal} {\bibinfo  {journal} {Soft Matter}\ }\textbf {\bibinfo {volume} {15}},\ \bibinfo {pages} {5810} (\bibinfo {year} {2019})}\BibitemShut {NoStop}%
\end{thebibliography}%

\pagebreak 
\newpage 

\onecolumngrid

\vspace{170mm}
\begin{center}
    {\textsc{ \Large{Supplementary Information} }}
\end{center}

\section{Details of the simulations}	
\subsection{Parameters}

The value of each parameter used in the simulations when not stated otherwise is given in Table~\ref{param}. Upon non-dimensionalisation, the unit length, time, strength and magnetic field are $4 \mu\textrm{m}$, $0.2 \si{.s}$,  $0.08 \si{.pN}$ and $1.14 \si{.mT}$ respectively.  

For computational feasibility, the number of bacteria is taken here to be at most $N_s=3000$. 

\begin{table}[h!]
\caption{\label{param} Parameters for the simulations, with experimental values~\cite{vincenti2019magnetotactic} and their non-dimensional equivalent, as well as the default values used in simulations.}
\centering
	\begin{tabular}{llcr}
	\toprule
		\textrm{Parameter}&
		\textrm{Experimental value}&
		\textrm{Non-dim.}&
		\textrm{Default} \\
		\midrule
		$ N_s $ &  $10^{17} \si{cells .m^{-3}} $ & $10^4$ & 500   \\
		$ n $ &  $N_s / 2 $   & $N_s / 2$  & $250$    \\
		$ B $ & $0.1 - 4 \si{.mT} $ & 0.11 - 4.5 & 5 \\  
		$ R $ & $20 - 120 \si{.\mu m}$  & 5 - 30 & 10  \\
		$ v_0 $ &$ 20- 40  \si{.\mu m. s^{-1}}$  &1 & 1 \\
		$ v_g $ & $ 3 \si{.\mu m. s^{-1}}$ & 0.15 & 0.15   \\
		$ \tilde{m} \mu  $ & $ 57 \si{.nN.\mu m^{-2} .mT^{-1} }$~\cite{Reufer_2014,Zahn_2017} & &  \\ 
		$ s $ & $ 0.57 \si{.pN}$~\cite{Hu_2015} & 7 & 7 \\
		$ \gamma $ & $  0.35  \si{.pN . \mu m}$~\cite{Hu_2015} & 1.1 & 0.4 \\
		$ \mu $ & $ 10^{-3} \si{.pN.s. \mu m^{-2}}$ &1 & 1  \\
		$ d $  & $ 4 \si{.\mu m}$~\cite{Hu_2015,Zahn_2017}&  1 &  1 \\
		$ D_r $ & $0.25 \si{. rad^2. s^{-1}}$\cite{Waisbord_2021} &  0.05 & 0 \\
		$ D_t $ & $200\si{. \mu m^2. s^{-1}}$~\cite{Ford_2007}  &  2.5 & 0.1 
		 \\
		\bottomrule
	\end{tabular}
\end{table}

\subsection{Components}
  Initially the $N_s$ bacteria are positioned uniformly in the sphere at $\bm{R} = (\bm{r}_i)$, with random unit orientations $\bm{P} = (\bm{p}_i)$~\cite{muller1959note}. 

The evolution of position and orientation is of swimmer $(i)$ is given by 
\begin{align}\label{eq:r_ev}
    \dot{\bm{r}}_i &= v_0 {\bm{p}}_i - v_g\hat{\bm{z}} + \bm{v}^{\text{im}}_{i}  + \bm{v}^{\text{ste}}_{\textrm{wall}\rightarrow i} +\sum_{j\neq i}  \left( \bm{v}^{\text{hyd}}_{j\rightarrow i}+\bm{v}^{\text{ste}}_{j\rightarrow i} \right) +\sqrt{2D_t}\, \bm{\xi}_t,\\
    \dot{\bm{p}}_i &= \Bigl[\bm{\omega}_i^B+\bm{\omega}^{\text{im}}_{i} +\sum_{j\neq i} \bm{\omega}^{\text{hyd}}_{j\rightarrow i}+\sqrt{2D_r}\, \bm{\xi}_r\Bigr] \times\bm{p}_i .\label{eq:p_ev}
\end{align}

With the contributions to the speed:
	\begin{itemize}[partopsep=2pt]
		\item constant swimming speed: $v_0 \bm{p}_i$,
		\item sedimentation speed  $v_g \hat{\bm{z}} $,
		\item hydrodynamic speed $\bm{v}^{\text{hyd}}_i $ from other bacteria $\bm{v}^{}_{j\rightarrow i}$ and all the images  $\bm{v}^{*}_{j\rightarrow i}$,
		\item steric hard core interactions with the bacteria $\bm{v} ^{\text{ste}}_{j\rightarrow i}$ with strength controlled by $d_{\text{ste}} = d$,
		\item steric interactions with the wall $\bm{v}^{\text{ste}}_{\textrm{wall}\rightarrow i}$ to ensure that $|\bm{r}_i| < R-d/2$ 
		\item noise with diffusion coefficient $D_t$ in translation, with with $\bm\xi_t$ a Gaussian white noise,
	\end{itemize}	
and to the rotation speed: 
	\begin{itemize}[partopsep=2pt]
		\item hydrodynamic reorientation from other bacteria $\bm{\omega}^{\text{hyd}}_{j\rightarrow i}$ and all the images $\bm{\omega}^{(h*)}_{j\rightarrow i}$,
		\item noise with diffusion coefficient $D_r$ in rotation, with with $\bm\xi_r$ a Gaussian white noise,
		\item magnetic reorientation.
	\end{itemize}

At each time step $\dd t$,  we compute the $(3 \times N_s)$ matrices for velocities $\bm{U} = (\bm{v}_i)$ and rotation velocity $\bm{\Omega} = (\bm{\omega}_i)$.  The  corresponding matrices for position and orientation,  $\bm{R}$ and $\bm{P}$ are then updated with a first order Euler method
	\begin{equation}
		\begin{aligned}
			\bm{R}(t+ \dd t) &  = \bm{R}(t )+ \dd t \, \bm{U} + \sqrt{D_t \dd t }\, \bm\xi  , \\ 
			\bm{P}(t+ \dd t) & = \bm{P}(t )+ \dd t \, \bm{\Omega}  \times  \bm{P}(t ) +\sqrt{D_r \dd t } \, \bm\xi \times \bm{P}(t )\quad  \textrm{and}  \quad  \bm{P}(t+ \dd t) = \bm{P}(t+ \dd t) / |\bm{P}(t+ \dd t)| .
		\end{aligned}
     \label{eqevol}
	\end{equation}

	\vspace{6pt} \noindent
	\textbf{Hydrodynamic interactions.}
	Each swimmer is a pair of Stokeslets $s$ and of rotlets $\gamma$ at $\pm d/2$ from its centre $\bm{x}_i$. We also assume that the swimmers are spherical and much smaller than the velocity gradient length scale,  so the advection velocity due to the flow of the swimmers in the no-slip sphere is equal to the flow speed at $\bm{x}_i$, namely 
	\begin{equation}
		\bm{v}^{\text{hyd}}_i = \sum_{j \neq i} \bm{v}_{j \rightarrow i} + \sum_{j } \bm{v}^{*}_{j \rightarrow i} .
	\end{equation}
The flow from swimmer $j \neq i$ is the sum of the flows due to the two point-forces and two point-torques, which reads
	\begin{equation}
		\begin{aligned}
			\bm{v}_{j \rightarrow i} = & \bm{v}^{(s)}\left(\bm{x}_i - (\bm{x}_j + d/2 \bm{p}_j) , s \bm{p}_j \right)  + \bm{v}^{(s)}\left(\bm{x}_i - (\bm{x}_j - d/2 \bm{p}_j) , - s \bm{p}_j \right) \\ & + \bm{v}^{(\gamma)}\left(\bm{x}_i - (\bm{x}_j + d/2 \bm{p}_j) , \gamma \bm{p}_j \right) + \bm{v}^{(\gamma)}\left(\bm{x}_i - (\bm{x}_j - d/2 \bm{p}_j) , - \gamma \bm{p}_j \right) ,
		\end{aligned}
	\end{equation}
The velocity from a single stokeslet and rotlet are  
\begin{equation}
		\bm{v}^{(s)} (\bm{x},\bm{s}) = \frac{1}{8 \pi \mu } \left[ \frac{\bm{s}}{r} + \frac{(\bm{s} \cdot \bm{x}) \bm{r}}{x^3} \right]
		\quad \textrm{and} \quad
		\bm{v}^{(\gamma)} (\bm{x},\bm{s}) = \frac{1}{8 \pi \mu }  \frac{\bm{\gamma} \times \bm{x}}{r^3} 
	\end{equation}
 respectively. 
The contribution from the hydrodynamic image is the sum of the individual images for the two stokeslets and the two rotlets. 
For the derivation, all the singularities are considered as the sum of an axisymmetric and a transverse component. 
The resulting flow for each term is found using results by~\citet{maul1994image} for the axisymmetric Stokeslet, \citet{shail1987note} for the transverse Stokeslet, \citet{chamolly2020stokes} for the axisymmetric rotlet and by~\citet{hackborn1986structure} for the transverse rotlet. Their expressions are detailed in section~\ref{secimages}.
	
The reorientation speed is obtained similarly,  assuming that the swimmers are spherical and much smaller than the length scale of the flow gradients
	\begin{equation}
		\bm{\omega}_i = \frac{1}{2} \left[\sum_{j \neq i} \bm{\omega}_{j \rightarrow i} + \sum_{j } \bm{\omega}^*_{j \rightarrow i}  \right],
	\end{equation}
with the contribution from a given swimmer again stemming from the sum of the four hydrodynamic singularities
	\begin{equation}
		\begin{aligned}
			\bm{\omega}_{j \rightarrow i} = & \bm{\omega}^{(s)}\left(\bm{x}_i - (\bm{x}_j + d/2 \bm{p}_j) , s \bm{p}_j \right)  + \bm{\omega}^{(s)}\left(\bm{x}_i - (\bm{x}_j - d/2 \bm{e}_j) , - s \bm{p}_j \right) \\ & + \bm{\omega}^{(\gamma)}\left(\bm{x}_i - (\bm{x}_j + d/2 \bm{e}_j) , r \bm{p}_j \right)  + \bm{\omega}^{(\gamma)}\left(\bm{x}_i - (\bm{x}_j - d/2 \bm{p}_j) , - r \bm{p}_j \right) .
		\end{aligned}
	\end{equation}
The vorticities from the singularities are 
	\begin{equation}
		\bm{\omega}^{(s)} (\bm{x},\bm{s}) =  \frac{2}{8 \pi \mu }  \frac{\bm{s} \times \bm{x}}{r^3} 
		\quad \textrm{and} \quad
		\bm{\omega}^{(\gamma)} (\bm{x},\bm{s}) = \frac{1}{8 \pi \mu } \left[ - \frac{ \bm{\gamma}}{r^3} + \frac{3 (\bm{\gamma} \cdot \bm{x}) \bm{x}}{r^5} \right] .
	\end{equation}
As we did for the velocities above, we additionally derive exactly and include the image vorticities for each hydrodynamic singularity, with the expressions given in section~\ref{secimages}.

The solutions that we use for the hydrodynamic velocities and vorticities are singular at $\bm{x}_j$. 
To ensure that the local values of the flow velocities induced by swimming remain physically relevant close to a particle, we, therefore, set maximum values for the local velocity and vorticity generated by a swimmer. 
Measurement of the flow near a swimming \textit{E. coli} by  \citet{drescher2011fluid} and detailed simulations~\cite{Hu_2015} support the idea of a threshold velocity for the flow in the near field.  We chose the threshold to be the maximum of the norm of the velocities and vorticities due to the pair of Stokeslets and rotlets, respectively,  at a distance of $d_{\text{min}}$ from the centre $\bm{x}_j$ of the swimmer.  In our simulations, we take $d_{\text{min}} = 1$. 
In the absence of a tractable regularisation function for stokeslets and rotlets inside a sphere,  $d_{\text{min}}$ acts as a proxy for regularisation as it prevents the velocities and vorticities from taking singular values.

The form of the hydrodynamic interactions used here is relevant for dilute suspensions.  As confined magnetotactic bacteria tend to accumulate against boundaries, we expect this assumption not to hold in some regions of the sphere for high numbers of swimmers,  comparable to the experimental concentration.  In such high-density regions, we would expect the flow from the swimmers to be screened and hence weaker than in our simulations.  We do not expect this restriction to affect the qualitative behaviour of the system, but the quantitative values of the simulated flows are likely to exceed experimental values. A possibility to get quantitative comparisons with experimental data could then be to rescale them to account for this screening.

	\vspace{6pt} \noindent \textbf{Steric interactions.}
To avoid overlaps between the swimmers, we include a steric interaction between swimmers that are less than a radius away from each other, of the form 
	\begin{equation}
		\bm{v}^{(ste)} = - \sum_{j \neq i} \delta_{ | \bm{x}_i - \bm{x}_j| < d} \left( \frac{ d_{\text{ste}}}{| \bm{x}_i - \bm{x}_j|} \right)^6  \frac{1 }{| \bm{x}_i - \bm{x}_j|} ( \bm{x}_i - \bm{x}_j ) .
	\end{equation}  
The steric repulsion decays as $1/r^6$ and its strength is set by the typical lengthscale $d_{\text{ste}}$, with a default value of $0.2$.

In addition,  after each time step, we ensure that the swimmers stay inside the drop and at a minimal distance $d/2$ of the boundaries.

	\vspace{6pt} \noindent \textbf{Sedimentation.}
The constant downward sedimentation speed $\bm{v}_g$ is obtained by considering the sedimentation of a sphere of diameter $d$ in the bulk of a viscous fluid. This gives $v_g = 2 (d/2)^2 g \Delta \rho / (9 \mu) \approx 3 \si{.\mu m .s^{-1}}$~\cite{guazzelli2011physical}.

	\vspace{6pt} \noindent \textbf{Magnetic reorientation.}
The magnetic torque reads
	\begin{equation}
		\bm{\omega}_m = \epsilon B \frac{m}{\zeta_s} \bm{p}_i \times \hat{\bm{y}} ,
	\end{equation}
	where $m$ is the magnetic moment and $\zeta_s$ is the drag on the swimmer for reorientation orthogonal to its long axis. Both were measured experimentally~\cite{Reufer_2014,Zahn_2017}, and we define $ \tilde{m} = m/\zeta_s $. 
  
We ensure that the magnetic field causes a reorientation at most aligned to the field axis $\hat{\bm{y}}$.  If overshooting occurs and $ |B \tilde{m}  \bm{p}_i \times \hat{\bm{y}} | > \arccos \left(\bm{p}_i \cdot \hat{\bm{y}} \right)$, we set instead that $\bm{p}_i (t+ \dd t) = \epsilon \hat{\bm{y}}$, so that the swimmer is aligned with the field and oriented towards its preferred pole.

	\vspace{6pt} \noindent \textbf{Noise.}
	We also include noise, both in translation and in orientation, with diffusion coefficient  $D_t$ and $D_r$  chosen to be independent, as shown in Eq.~\eqref{eqevol}~\cite{Zottl_2019}.

\subsection{Implementation}
There are also two parameters in our model that are not measured experimentally, the distance which sets the maximal hydrodynamic speed $d_{\text{min}} / d$ and that for steric interactions $d_{\text{ste}} /d$. 
	
In our simulations, we can tune macroscopic parameters, the number of swimmers, the magnetic field and the drop radius, and microscopic ones, the strength of gravity, the strength of the forces and torques due to the propulsion,  the strength of steric interactions and the noise. 
We use a dimensionless time step $\dd t = 0.05$ and an explicit Euler scheme to integrate the equations of motion.

\vspace{140pt}

\section{Influence of the microscopic parameters}

\subsection{Collective effects with one or two populations and achiral swimmers}
\begin{figure}[h!]
\centering
		\includegraphics[width = .85\columnwidth]{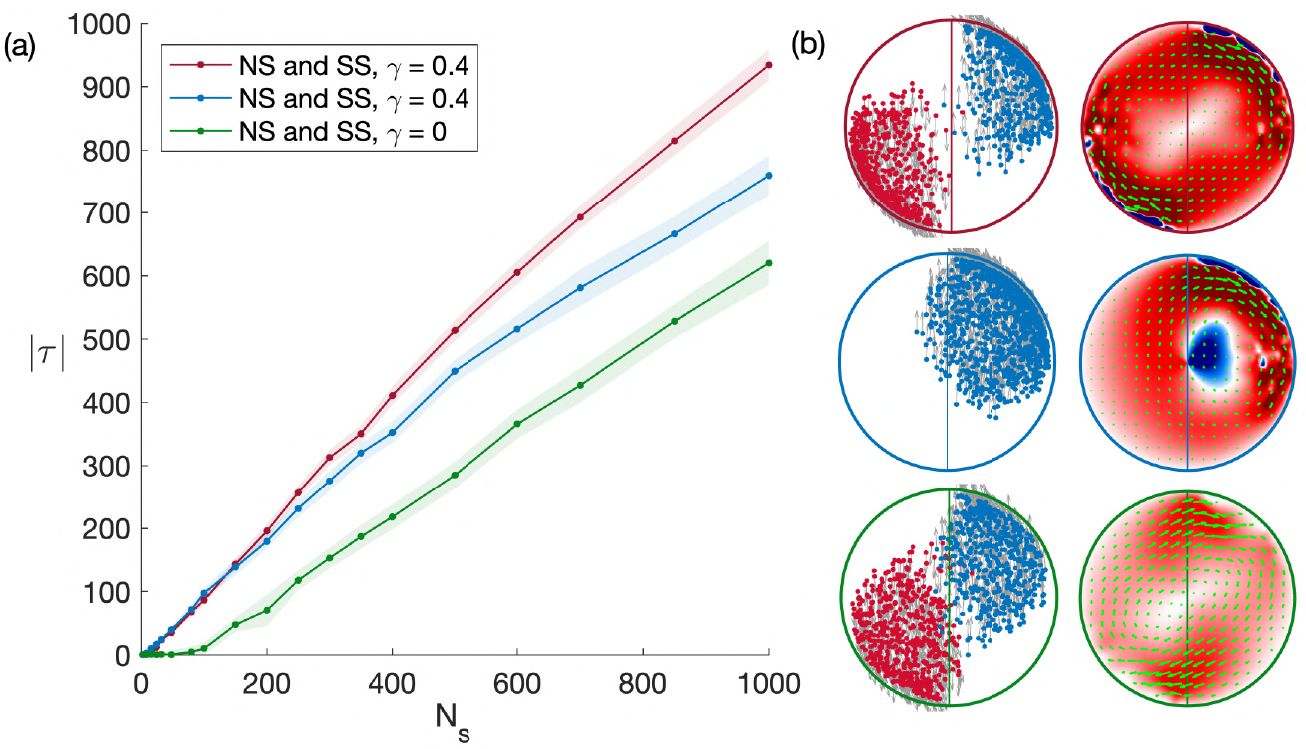}
		\caption[Chiral and achiral symmetry breaking for one or two populations.]{\label{Nscomp} Symmetry-breaking for two (red, top) or one (blue, centre) populations of chiral swimmers, and two populations of achiral swimmers (green, bottom) at $B = 5$.  (a) Absolute value of the torque on the drop for $N_s$ swimmers, chiral or not, split into two populations or not, with the shaded region showing the standard deviation in time. (b) Position of the swimmers and equatorial flow for $N_s = 1000$. With a SS and a NS population, a symmetry breaking occurs and a torque arises for both achiral and chiral swimmers, but the chiral mechanism is both stronger and more robust at high concentrations.  It also has a CW direction, while the achiral torque can be either CW or CCW. A torque is also created for a single chiral population but saturates for higher numbers of swimmers.  
		}
\end{figure}

We run simulations for different numbers of swimmers when taking into account one (NS) or two (both NS and SS) populations of swimmers in Fig.~\ref{Nscomp}. 

For a single NS population of achiral ($\gamma = 0$) swimmers, no vortex forms. If both NS and SS swimmers are present on the other hand, a vortex forms above a threshold number of swimmers. This vortex has no preferred CW-CCW direction. 
It also generates a torque on the droplet, which is plotted in Fig.~\ref{Nscomp} for varying concentrations of swimmers. 

A vortex forms for a single NS population of chiral swimmers, but the resulting torque grows slower with $N_s$ than for two populations. 

\subsubsection{Interplay of force-based propulsion, gravity and chirality}

\begin{figure}[b]
\centering
		\includegraphics[width = .85\columnwidth]{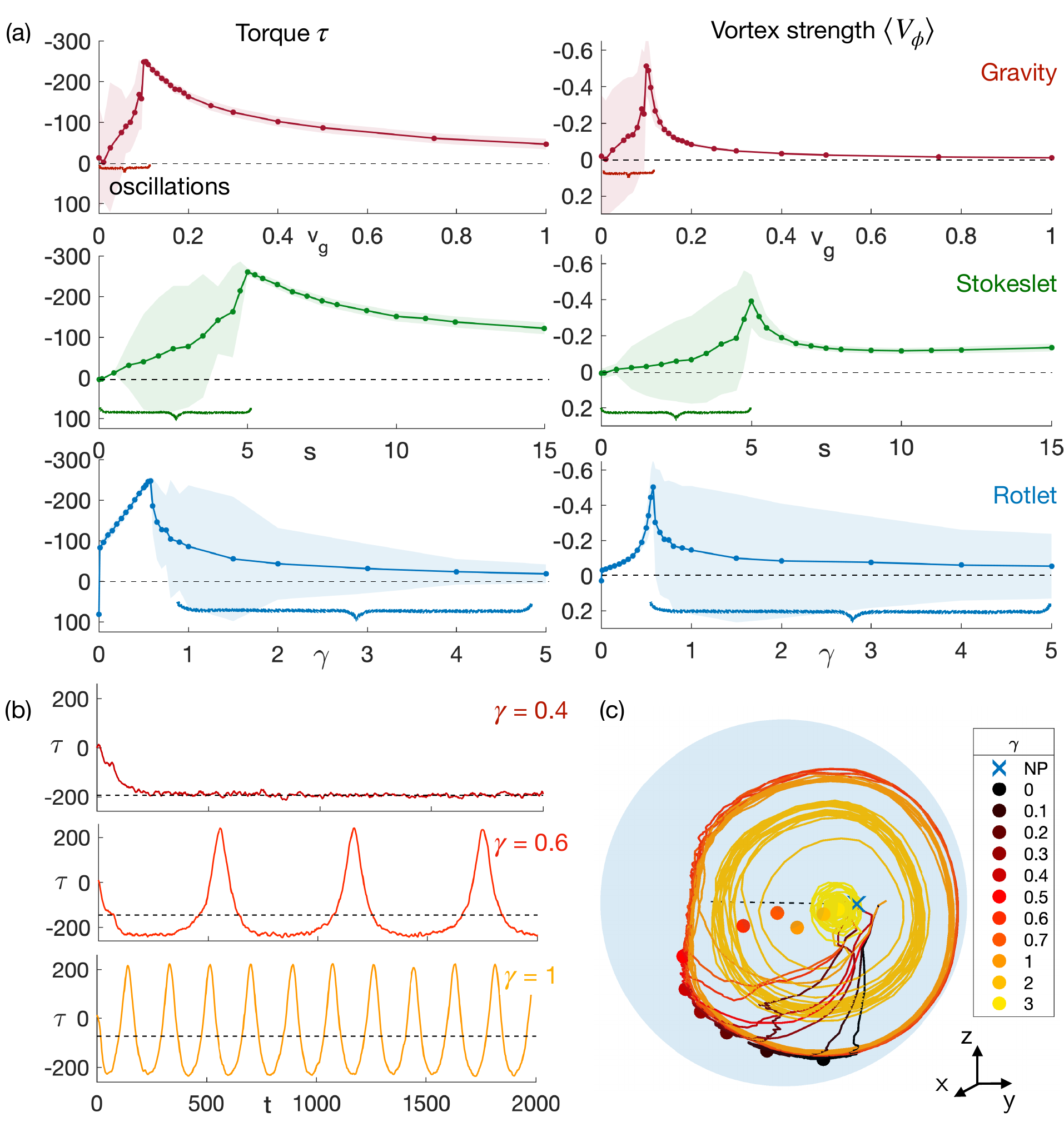}
		\caption[Variation of the vortex and the torque with the microscopic parameters]{\label{figvsg} Changing the microscopic parameters and transition to oscillations. (a) Torque on the drop (left) and vortex strength (right) when varying gravity $v_g$, Stokeslets $s$ and rotlets $\gamma$. In each case, there is an optimal value past which chirality becomes dominant and the long-term state is oscillatory, which leads to an increase in standard deviation outlined with an accolade.  (b) Time variation of the torque on the drop due to 200 NS swimmers, for three values of chirality, $\gamma = 0.4, 0.6$ and $1$, with the late time mean torque shown as a dashed line.  For $\gamma = 0.6, 1$, the torque displays oscillations and changes direction.  (c) Trajectory of the centre of a group of 200 NS swimmers with time for different values of $\gamma$, and onset of oscillations.  The average late-time position of the cluster is shown as a dot. }
\end{figure}

We vary the Stokeslet strength $s$, rotlet strength $\gamma$ and gravity $v_g$ while keeping constant the other two parameters, and plot the corresponding torque on the droplet and vortex strength in Fig.~\ref{figvsg}a. 
In all three cases, there is an optimal value for the collective motion which maximises both the average azimuthal velocity and the torque.  
When the system is less chiral than this, so for lower $\gamma$ or conversely higher $v_g$ or $s$, both vortex strength and torque decrease steadily. 
On the other hand, for very chiral swimmers,  the stationary state of the system disappears.  
At long times, both clusters then undergo periodic trajectories around their pole, not always in phase,  and the vortex flow alternates between CW and CCW directions.  These oscillations for very chiral swimmers also occur with a single population and are plotted in Fig.~\ref{figvsg}c. As the cluster rotates around the pole, the resulting vortex and torque on the drop also change sign (Fig.~\ref{figvsg}d) while remaining CW on average. This leads to the decrease in the mean late time value and the large standard deviation for high relative chirality in Fig.~\ref{figvsg}a. 
The exact value for the transition depends strongly on the details of the model, including in the near field. 
Importantly, however, this oscillating state in our system occurs for physically relevant values of the microscopic parameters.

	\subsection{Robustness to noise}
\begin{figure}[h!]
	\centering
		\includegraphics[width =   \columnwidth]{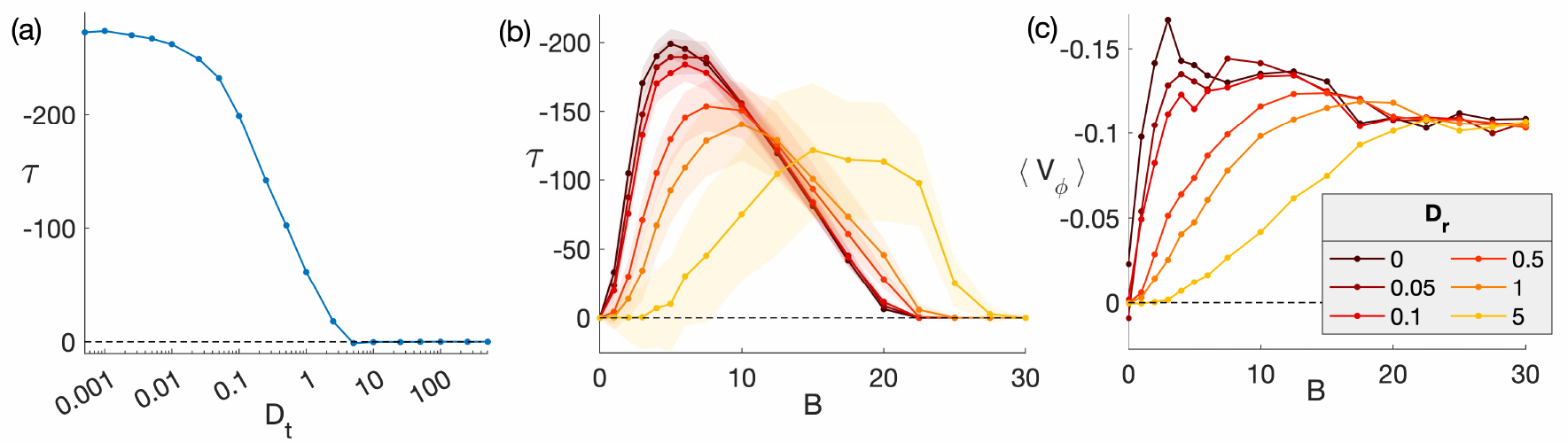}
		\caption[Robusteness of the system to translational and rotational noise]{\label{noise} (a) Variation of the strength of the torque $\tau$ on the drop with the translational noise strength $D_t$.  (b) Variation of the torque $\tau$ on the drop with the magnetic field $B$  for different rotational noise strength $D_r$. The shaded areas show the standard deviation in time.  The value of the rotational noise in the bulk is  $D_r \approx 0.05$ in dimensionless units~\cite{Waisbord_2021}. (c) The time-average of the azimuthal velocity $\langle V_\phi \rangle$ averaged over $r$ depending on the field $B$ for the same values of $D_r$. }
\end{figure}

	We test the robustness of our system to noise both in orientation and translation,  each assumed independent from the other, in Fig.~\ref{noise}.  

A strong translational noise reduces the strength of collective effects, and in particular, the torque on the system $\tau$, as shown in Fig.~\ref{noise}.
Increasing the rotational noise $D_r$ tends to delay the apparition of the vortex to higher values of the magnetic field, as shown in Fig.~\ref{noise}c.  The peak value for the vortex strength is also smoothed.  At high fields,  increasing $D_r$ has no effect on the system as the torque from the magnetic fields dominates. 

When considering values corresponding to the diffusion of a single magnetotactic bacteria in the bulk $D_r \sim 0.25 \si{rad. s^{-1}}$~\cite{Waisbord_2021}, the influence of the rotational noise on the transition is small.  
However, in the suspension, we would expect contacts between swimmers to be an important source of noise in the system. While the corresponding values of $D_t$ and $D_r$ are hard to estimate, we show in Fig.~\ref{noise} the effect of increased noise up to strengths much higher than the bulk ones. 
Overall, the qualitative properties of the system are very robust to noise,  but we would expect the active noise in a suspension to affect quantitative values such as the strength of the field that maximises the rotation.  
	
\section{Images of force and torque dipoles inside a no-slip sphere}
\label{secimages}
	
	Here we derive expressions for the velocity and vorticity fields due to axisymmetric and transverse Stokeslets and rotlets inside of a rigid sphere. The derivation is based heavily on previous work carried out by previous authors, namely \cite{maul1994image} for the axisymmetric Stokeslet, \cite{shail1987note} for the transverse Stokeslet, \cite{chamolly2020stokes} for the axisymmetric rotlet and \cite{hackborn1986structure} for the transverse rotlet. 
	
	Throughout this section we shall assume that the sphere is centred at the origin and has unit radius, that the viscosity $\mu=\left(8\pi\right)^{-1}$ and that the singularity is located at $\bm{x}_1=\{0,0,c\}$ in Cartesian coordinates with $0<c<1$ and has unit magnitude. The solution for more general parameters may then be obtained by applying rotations and scalings in conventional fashion.
	
	Furthermore, we define the spherical polar coordinates
	\begin{align}
		r = \sqrt{x^2+y^2+z^2}, \quad \theta = \cos^{-1}\left(z/r\right), \quad \phi=\tan^{-1}\left(y/x\right)
	\end{align}	 
	and the auxiliary quantities
	\begin{align}
		\bm{x}_2 &= \{0,0,c^{-1}\},\\
		\bm{r}_1 &= \bm{x}-\bm{x}_1, \\
		\bm{r}_2 &= \bm{x}-\bm{x}_2, \\
		r_1 &= |\bm{x}-\bm{x}_1| = \sqrt{r^2 + c^2-2rc\cos\theta},\\
		r_2 &= |\bm{x}-\bm{x}_2| = \sqrt{r^2 + c^{-2}-2rc^{-1}\cos\theta}.
	\end{align}
	
	In each case, the solution satisfies the boundary condition $\bm{u}=\bm{0}$ on the surface of the unit sphere and the Stokes equations within, except for the desired kind of singular behaviour at $\bm{x}_1$.

	\subsection{Axisymmetric Stokeslet}
	
	We consider a point force directed in the positive $z$-direction, $\bm{e}=\{0,0,1\}$. \cite{maul1994image} give the solution in terms of singularities expressed as the Oseen tensor and its derivatives. Specifically,
	\begin{align}
		\bm{u}\left(\bm{x}\right)=\bm{\mathcal{G}}\left(\bm{r}_1\right)\cdot\bm{e} + \frac{1 - 3 c^2}{2 c^3} \bm{\mathcal{G}}\left(\bm{r}_2\right)\cdot\bm{e} - \frac{1 - c^2}{c^4}\left(\bm{e}\cdot\bm{\nabla}\right)\bm{\mathcal{G}}\left(\bm{r}_2\right)\cdot\bm{e} - \frac{\left(1 - c^2\right)^2}{4 c^5} \nabla^2\bm{\mathcal{G}}\left(\bm{r}_2\right)\cdot\bm{e} ,
	\end{align}
	where the Oseen tensor is given by
	\begin{align}
		\bm{\mathcal{G}}\left(\bm{r}\right)=\frac{\bm{I}}{|\bm{r}|}+\frac{\bm{r}\bm{r}}{|\bm{r}|^3}.
	\end{align}
	Taking components and the curl respectively it is then straightforward (with appropriate software) to find components of the velocity and vorticity in spherical polar coordinates as follows.
		\begin{dmath}
			u_r = \frac{1}{2} \left(2 \cos \theta \left(\frac{(c-r \cos
				\theta)^2}{r_1^3}+\frac{1}{r_1}+\frac{\left(1-3 c^2\right)
				\left(\left(\frac{1}{c}-r \cos \theta\right)^2+r_2^2\right)}{2 c^3
				r_2^3}+\frac{\left(-1+c^2\right)^2 \left(-3 r^2+3 r^2 \cos (2 \theta)+4
				r_2^2\right)}{4 c^5 r_2^5}+\frac{\left(-1+c^2\right) (-1+c r
				\cos \theta) \left(-3-3 c r \cos \theta (-2+c r \cos \theta)+c^2 r_2^2\right)}{c^7
				r_2^5}\right)+r \cos ^2\phi \left(-\frac{2 (c-r \cos
				\theta)}{r_1^3}+\frac{3-3 c^2 r_2^2+c^4 \left(-3+5
				r_2^2\right)+c r \cos \theta \left(-9-6 c \left(-1+c^2\right) r \cos \theta-3
				c^4 \left(-1+r_2^2\right)+c^2 \left(6+r_2^2\right)\right)}{c^6
				r_2^5}\right) \sin ^2\theta+r \left(-\frac{2 (c-r \cos
				\theta)}{r_1^3}+\frac{3-3 c^2 r_2^2+c^4 \left(-3+5
				r_2^2\right)+c r \cos \theta \left(-9-6 c \left(-1+c^2\right) r \cos \theta-3
				c^4 \left(-1+r_2^2\right)+c^2 \left(6+r_2^2\right)\right)}{c^6
				r_2^5}\right) \sin ^2\phi \sin ^2\theta\right),
		\end{dmath}
		\begin{dmath}
			u_\theta = \frac{\left(-4 c^9 r_2^5-4 c^7 r_1^2
				r_2^5+r_1^3 \left(12+c^2 \left(-12+3 \left(3-5 c^2+2 c^4\right)
				r^2+2 r_2^2 \left(-5+c^2 \left(9-2 c^2+\left(-1+3 c^2\right)
				r_2^2\right)\right)\right)\right)\right) \sin \theta}{4 c^7 r_1^3
				r_2^5}+\frac{r \left(\left(2 c^7
				r_2^5+r_1^3 \left(-15+c^2 \left(18+r_2^2-3 c^2
				\left(1+r_2^2\right)\right)\right)\right) \sin (2 \theta)-3 c
				\left(-1+c^2\right) r r_1^3 \sin (3 \theta)\right)}{4 c^6 r_1^3
				r_2^5},
		\end{dmath}
		\begin{dmath}
			u_\phi=0,
		\end{dmath}
		\begin{dmath}
			\omega_r=0,
		\end{dmath}
		\begin{dmath}
			\omega_\theta=0,
		\end{dmath}
		\begin{dmath}
			\omega_\phi=\frac{r \left(\left(4 c^7 r_2^7+r_1^3 \left(-15 \left(-1+c^2\right)^2
				\left(1+c^2 r^2\right)+3 c^2 \left(1-6 c^2+5 c^4\right) r_2^2+2 c^4 \left(1-3
				c^2\right) r_2^4\right)\right) \sin \theta\right)}{2 c^7 r_1^3
				r_2^7}-\frac{3  \left(-1+c^2\right) r^2 r_1^3
				\left(5+c^2 \left(-5+2 r_2^2\right)\right) \sin (2 \theta)}{2 c^6 r_1^3
				r_2^7}.
		\end{dmath}

	\subsection{Transverse Stokeslet}
	For the transverse Stokeslet we assume that the unit force is oriented in the positive $x$-direction, $\bm{e}=\{1,0,0\}$. The solution by \cite{shail1987note} is given in terms of hydrodynamic scalar potentials $\psi(r,\theta)$ and $\chi(r,\theta)$ from which the flow components are calculated as
	\begin{align}
		u_r &= -\frac{\cos\phi}{r^2} \frac{\partial }{\partial \theta}\left(\frac{1}{\sin \theta}\frac{\partial \psi}{\partial\theta}\right),\\
		u_\theta &= \frac{\cos \phi}{r} \frac{\partial }{\partial \theta}\left(\frac{1}{\sin \theta}\frac{\partial \psi}{\partial r}\right)+\frac{\cos \phi}{r \sin ^2\theta}\chi,\\
		u_\phi &= -\frac{\sin \phi }{r \sin ^2\theta}\frac{\partial \psi }{\partial r}-\frac{\sin \phi}{r}\frac{\partial }{\partial \theta}\left(\frac{\chi}{\sin \theta}\right).
	\end{align}
	The potentials themselves are given as follows (we note that our definition of $r_2$ differs from \cite{shail1987note}).
	\begin{dgroup}
		\begin{dmath}
			\psi = \frac{(r \cos \theta-c) r_1}{c}+\frac{(r-1)^2 \left(\left(r \left(1-c^2\right)-2
				c^2\right) \cos \theta-2 c\right)}{2 c}+\frac{4 c r^2 \cos
				^2\theta-\left(\left(c^2+1\right) r^2+\left(5 c^2+1\right)\right) r \cos \theta+c
				\left(\left(1-c^2\right) r^4+\left(3 c^2-1\right) r^2+2\right)}{2 c^2 r_2},
		\end{dmath}
		\begin{dmath}
			\chi = \frac{2 (r_1-c r_2-(r-1) (c \cos \theta+1))}{c}.
		\end{dmath}
	\end{dgroup}
	From this, we find the velocity components as
	\begin{dgroup}
		\begin{dmath}
			u_r = \frac{1}{2} \cos \phi \left(-\frac{2 c (c-r \cos\theta)}{r_1^3}+\frac{4}{r_1}+\frac{1}{c^4 r_2^5}\left(3 r \left(1+r^2+c^2\left(5+r^2\right)\right) \cos \theta+3 c \left(-2-\left(1+3 c^2\right)r^2+\left(-1+c^2\right) r^4-2 r^2 \cos (2\theta)\right)+2 c \left(1+r^2+c^2\left(5+r^2\right)-8 c r \cos \theta\right) r_2^2-8 c^3 r_2^4\right)\right) \sin \theta,
		\end{dmath}
		\begin{dmath}
			u_\theta= \frac{\cos \phi \csc ^2\theta}{4 c^2 r r_2^3} \left(-4-2 \left(3+11 c^2\right) r^2+4 \left(-2+c^2\right)
			r^4+\frac{r}{c}\left(\left(3+4 r^2+c^2 \left(19+6 \left(2+c^2\right) r^2-2
			\left(-1+c^2\right) r^4\right)\right) \cos \theta-2 c \left(-1+c^2\right) r \left(-3+2
			r^2\right) \cos (2\theta)-\left(1+2 r^2+c^2 \left(5+4 r^2\right)\right) \cos (3\theta)+4 c r
			\cos (4 t)\right)+2 \left(1+5 c^2+3 \left(1+c^2\right) r^2+2 c r \cos \theta
			\left(-5-3 c^2+2 \left(-1+c^2\right) r^2+2 \cos (2\theta)\right)\right)
			r_2^2-8 c^2 r_2^4+\frac{3 r}{c^2 r_2^2} \left(-2 c \left(2+\left(3+8
			c^2\right) r^2+2 r^4\right) \cos \theta+r \left(1+r^2+c^2 \left(9+r^2 \left(3+2 r^2-2
			c^2 \left(-3+r^2\right)\right)\right)+\left(1+5 c^2\right) \left(1+r^2\right) \cos (2
			t)-2 c r \cos (3\theta)\right)\right) \sin ^2\theta+\frac{2 c
				r_2^3}{r_1^3} \left(r_1^2 \left(-2 \left(c^2+r^2\right)+5 c r \cos
			\theta-c r \cos (3\theta)+3 \left(-1+c^2\right) \left(-1+r^2\right) r_1+2
			r_1^2\right)+c r \left(-2 \left(c^2+r^2\right) \cos \theta+c r (3+\cos (2
			t))\right) \sin ^2\theta\right)\right),
		\end{dmath}
		\begin{dmath}
			u_\phi =-\frac{\csc ^2\theta}{2
				c^2 r r_2^2} \left(\frac{(c r-\cos \theta)}{c
				r_2} \left(r \left(1+r^2+c^2
			\left(5+r^2\right)\right) \cos \theta+c \left(-2-\left(1+3 c^2\right)
			r^2+\left(-1+c^2\right) r^4-2 r^2 \cos (2\theta)\right)\right)-\left(-6 c^3 r+4 c \left(-1+c^2\right) r^3+\left(1+5 c^2+3
			\left(1+c^2\right) r^2\right) \cos \theta-6 c r \cos (2\theta)\right) r_2-\frac{c}{r_1}
			\left(-2 \left(c^2+r^2\right) \cos \theta+c r (1+3 \cos (2\theta))+\cos \theta r_1
			\left(3 \left(-1+c^2\right) \left(-1+r^2\right)+2 r_1\right)\right)
			r_2^2+4 c^2 \cos \theta r_2^3\right) \sin \phi.
		\end{dmath}
	\end{dgroup}
	
	The vorticity components are omitted due to their length.
	
	\subsection{Axisymmetric rotlet}
	For the axisymmetric rotlet we again consider an orientation in the positive $z$-direction, $\bm{e}=\{0,0,1\}$, corresponding to an anticlockwise rotation when viewed from $z=\infty$. \cite{chamolly2020stokes} found that the solution can be written with the aid of just one additional image rotlet and reads
	\begin{align}
		u_r &= u_\theta = 0,\\
		u_\phi  &= r\left(\frac{1}{r_1^3}-\frac{1}{c^3r_2^3}\right)\sin\theta.
	\end{align}
	The vorticity is given by
	\begin{align}
		\omega_r &= \frac{2 \cos \theta}{r_1^3}-\frac{3 c r \sin ^2\theta}{r_1^5}+\frac{-2 c
			\cos \theta r_2^2+3 r \sin ^2\theta}{c^4 r_2^5},\\
		\omega_\theta &= \left(\frac{3 r (r-c \cos \theta)}{r_1^5}-\frac{2}{r_1^3}+\frac{3 r
			(-c r+\cos \theta)+2 c r_2^2}{c^4 r_2^5}\right) \sin \theta,\\
		\omega_\phi &= 0.
	\end{align}
	
	\subsection{Transverse rotlet}
	For the transverse rotlet, we assume the orientation is $\bm{e}=\{0,1,0\}$, so that the rotation is in the $x$-$z$-plane. The solution found by \cite{hackborn1986structure} is again in terms of potentials, much like the transverse Stokeslet found by \cite{shail1987note}, and the potentials are
	\begin{align}
		\psi &= -\frac{r_1}{c}+\frac{r}{c}-\cos \theta+\frac{r}{c r_2} \left(1-r^2\right) (c r-\cos
		\theta)+\frac{1}{2} r \left(3-r^2\right) \cos \theta+\frac{1}{2}
		\left(1+r^2\right) \left(r_2-\frac{1}{c}\right),\\
		\chi &= \frac{1}{r_1}-\frac{r \cos \theta}{c r_1}+\frac{\cos \theta}{c}-\frac{r}{c^2}
		\left(\frac{c r-\cos \theta}{r_2}+c \cos \theta\right).
	\end{align}
	Explicit expressions for the velocity and vorticity are again omitted here due to their extreme length.

\end{document}